\documentclass[journal,draftcls, onecolumn]{IEEEtran}
\usepackage[dvipdfmx]{graphicx}
\usepackage{amsmath}
\usepackage{amssymb}
\usepackage{amsfonts}
\usepackage{pifont}
\usepackage{mathrsfs}
\usepackage{theorem}
\usepackage{color}
\usepackage{float}
\usepackage{subfig}
\DeclareMathAlphabet{\bm}{OML}{cmm}{b}{it}
\theorembodyfont{\rmfamily}
\newtheorem{theorem}{Theorem}
\newtheorem{lemma}{Lemma}
\newtheorem{definition}{Definition}
\newtheorem{corollary}{Corollary}
\newtheorem{remark}{Remark}
\newtheorem{proposition}{Proposition}
\newtheorem{example}{Example}

\newcommand{\qed}{\hfill \IEEEQED}

\newcommand{\markov}{ - \!\!\circ\!\! - }

\newcommand{\bol}[1]{\mathbf{#1}}
\newcommand{\rom}[1]{\mathrm{#1}}
\newcommand{\san}[1]{\mathsf{#1}}

\newcommand{\Pe}{\rom{P}_{\rom{e}}}

\newcommand{\textchange}[1]{#1}

\begin{document}

\title{
A Classification of Functions 
in Multiterminal Distributed Computing\thanks{A part of this paper is submitted to IEEE International Symposium on Information Theory 2018.}
}

\author{Shun Watanabe\thanks{The work of S.~Watanabe is supported
in part by JSPS KAKENHI Grant Number 16H06091.}}


\maketitle
\begin{abstract}
In the distributed function computation problem, dichotomy theorems, initiated by Han-Kobayashi, seek to classify functions
by whether the rate regions for function computation improve on the Slepian-Wolf regions or not.
In this paper, we develop a general approach to derive converse bounds on the distributed function computation problem.
By using this approach, we recover the sufficiency part, i.e. the conditions such that the Slepian-Wolf regions become optimal,
of the known dichotomy theorems in
the two-terminal distributed computing. Furthermore, we derive an improved sufficient condition on the dichotomy theorem in
the multiterminal distributed computing for the class of i.i.d. sources with positivity condition. Finally, we derive the matching sufficient
and necessary condition on the dichotomy theorem in
the multiterminal distributed computing for the class of smooth sources.
\end{abstract}

\section{Introduction} \label{sec:introduction}

\begin{figure}[t]
\centering{
\includegraphics[width=0.45\textwidth, bb=0 0 301 164]{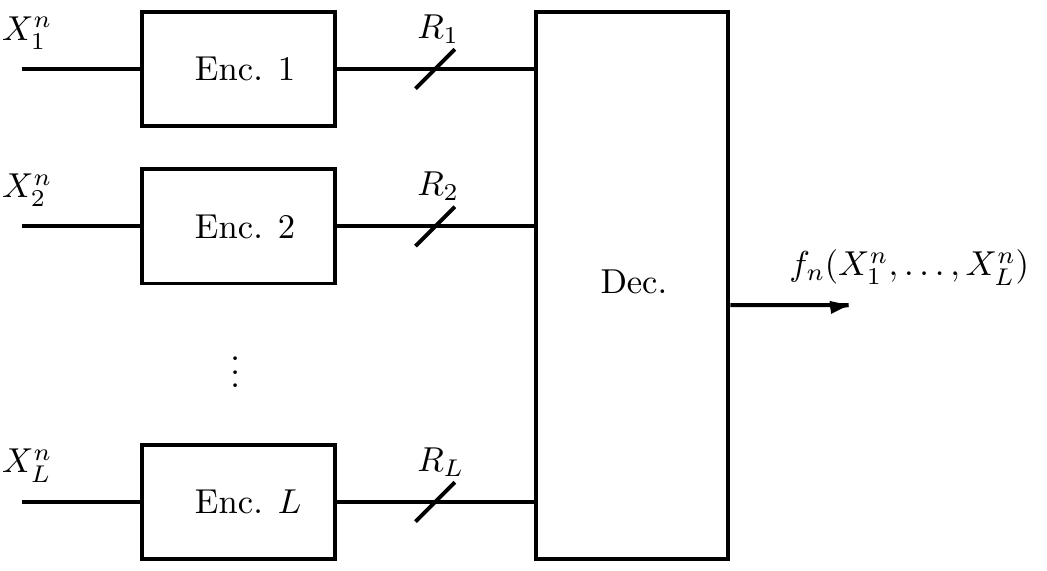}
\caption{A description of multiterminal distributed computing.}
\label{Fig:system}
}
\end{figure}

The distributed function computing is one of the most basic but difficult problems in network information theory.
In this problem, correlated sources are observed at $L$ terminals, and separately encoded messages are sent 
to the decoder so that a function value of the sources can be computed at the decoder; see Fig.~\ref{Fig:system}.
A naive scheme to compute a function is to first reproduce the entire source at the decoder and then to compute the function value.
Thus, it is apparent that the Slepian-Wolf (SW) region \cite{slepian:73, cover} is an inner bound on the achievable rate region for the function computing
problem. However, since the entire source needs not be reproduced at the decoder, the SW region can be improved in general.
Then, we are interested in under what conditions the SW region can be improved. For instance, it is well known that,
when the number of terminals is two and the function to be computed is the modulo-sum of the binary double symmetric source, 
the K\"orner-Marton (KM) coding improves upon the SW region \cite{korner:79}.\footnote{More precisely, it was shown in \cite{korner:79}
that the structured coding improves upon
the two-helper extension of the Wyner-Ahlswede-K\"orner (WAK) region \cite{wyner:75c, ahlswede:75}. 
In a similar manner as \cite[Lemma 1 and Lemma 2]{HanKob87}, 
it can be verified that the two-helper extension of the WAK region coincides with the SW region for the modulo-sum function. It should be also noted that
Ahlswede-Han (AH) showed a hybrid use of the structured coding and the two-helper WAK scheme further improves upon 
the convex hull of the SW region and the KM region \cite{AhlHan:83}. However, as is conjectured in \cite{SefGohRez:15},
the minimum sum-rate
of the AH region may not improve on the minimum sum-rate of the convex hull of the SW region and the KM region for the modulo-sum function.}

In \cite{HanKob87}, Han-Kobayashi initiated the study of classifying symbol-wise functions by whether 
the SW region can be improved or not.\footnote{The symbol-wise function means that, for a given function on a single observation space, we compute
the same copies of the function for a sequence of observations.} 
\textchange{When the number of terminals is two, they
completely characterized the condition of classification; the derived condition,
termed ``dichotomy theorem", only depends on the structure of functions in the following sense: if a given function satisfies the condition, then
the achievable rate region of computing that function coincides with the SW region for 
any independent and identically distributed (i.i.d.) sources
as long as the the positivity condition is satisfied;\footnote{The positivity condition
is the condition such that all symbols occur with positive probability.}  on the other hand, if a given function does not satisfy the condition,
then there exists an i.i.d. source satisfying positivity condition such that the SW region can be strictly improved.}
They also studied the classification problem for more than
two terminals, and derived a partial solution to the problem; however, the complete characterization remained as an open problem. 

\textchange{It should be noted that, when a given function violates the Han-Kobayashi condition,
the dichotomy theorem only claims existence of a source such that the SW region can be improved;
it is more challenging problem to decide if the SW region can be improved or not
for a given i.i.d. source with positivity condition. For instance, the modulo-sum function violates the Han-Kobayashi condition, and, as we mentioned above,  
the SW region can be improved for the binary double symmetric source. However, when the minimum of entropies of marginal distributions is smaller than
the entropy of the modulo-sum, the optimal region for computing the modulo-sum coincides with the SW region \cite[Problem 16.23(b)]{csiszar-korner:11}. As long as the author know,
it is an open problem to decide whether the optimal region for computing the modulo-sum coincides with the SW region or not for a given binary double source.}

So far, the above mentioned results are for i.i.d. sources. 
In \cite{KuzWat15}, Kuzuoka-Watanabe introduced a class of sources termed ``smooth sources".
This class of sources includes i.i.d. sources with positivity condition as a special case; but it also includes
sources with memory, such as Markov sources with positive transition matrices, and non-ergodic sources, such
as mixtures of i.i.d. sources with positivity condition. 
Then, they considered the function classification problem
for smooth sources and derived an alternative dichotomy theorem that is different from the one in \cite{HanKob87}.
More specifically, since the class of sources considered in \cite{KuzWat15} is broader than that considered in \cite{HanKob87},
the condition on functions in \cite{KuzWat15} is more strict than that in \cite{HanKob87}. 
\textchange{This difference stems from the fact that, even when a given function satisfies the Han-Kobayashi condition, 
the SW region may be improved for sources having memory, which signifies an importance of studying distributed function computing
beyond i.i.d. sources.}

\textchange{There are three motivations in this paper:}
\begin{enumerate}
\renewcommand{\theenumi}{\roman{enumi}}
\item \textchange{As we mentioned above, only a partial solution to the classification problem for more than two terminals was derived in \cite{HanKob87}.
Even for the simplest setting of three terminals with binary sources, there are some functions that cannot be classified by the conditions derived in \cite{HanKob87}.
In order to make a progress on this long-standing open problem, it is desirable to have a new approach to tackle this problem.}

\item \textchange{For the dichotomy theorems of two terminals, the sufficiency part\footnote{The sufficiency part of dichotomy theorems claim that 
there exist no codes that improve upon the SW region.} of Han-Kobayashi \cite{HanKob87} and Kuzuoka-Watanabe \cite{KuzWat15} were derived by completely different methods. 
The former is based on the single letter characterization; sources being i.i.d. is crucial, and it cannot be applied for general smooth sources.
On the other hand, the latter is based on the argument introduced by El Gamal \cite{ElGamal:83}; 
functions having certain structure called ``total sensitivity" is crucial. The condition of total sensitivity is more restrictive than the Han-Kobayashi condition, and
the proof method in \cite{KuzWat15} does not reproduce the dichotomy theorem of \cite{HanKob87} when sources are restricted to i.i.d. sources.
Thus, it is desirable to have an approach that provides both the dichotomy theorems in a unified manner;
currently, when a given function satisfies the Han-Kobayashi condition but not total sensitivity,
we cannot tell if the function computing region coincides with the SW region even for
a simple example of smooth sources such as mixtures of i.i.d. sources.
}

\item \textchange{For the two terminal distributed computing where the decoder has full side-information,
a method to derive converse bounds was proposed in \cite{KuzWat16}.  In that method, from the nature of distributed computing
and the structure of the function to be computed, termed {\em informative structure}, we identify information that is inevitably conveyed to the decoder.
Then, we derive a bound in terms of the optimal rate needed to send that information;
when the inevitably conveyed information is the encoder's source itself, we can conclude that the Slepian-Wolf rate is optimal.
In \cite{KuzWat16}, in addition to recovering some of known results in \cite{AhlCsi:81, ElGamal:83, OrlRoc:01}, 
the method was used to provide some novel results for the distributed computing with full side-information.
This method has a potential to resolve problems raised in the above mentioned two motivations. However, 
the method crucially relies on the fact that the decoder has full side-information,
and it does not apply to the case where all observations are encoded; in the conclusion of \cite{KuzWat16}, the authors left
such an extension as an important future research direction.
}
\end{enumerate}

\textchange{In this paper, we develop a general approach for showing 
dichotomy theorems, in particular, the sufficiency part.
The proposed approach is build upon the proof method based on informative structure proposed in \cite{KuzWat16}, but requires two new ideas described below. }

As we mentioned above, the proof method based on informative structure relies on the fact that the decoder has full side-information,
and it does not apply to the case where all observations are encoded. 
However, when the function induces conditional independence (CI) structure to the observed sources for some partition of the terminals,
then we can virtually decouple the entire coding system into multiple coding systems where
each decoder observe sources corresponding to parts of the partition as full side-information.
Then, we can apply the proof method based on informative structure to each decoupled coding system.

It should be noted that the CI structure and its relaxation have been known as crucial for proving converse in
multiterminal distributed computing or multiterminal rate-distortion problems
\cite{GelPin:79, berger:96, viswanathan:97, oohama:98, oohama:05, PraTseRam:04, wagner:08, wagner:08b, Oohama:06b, oohama:09b, TavVisWag:10, SefTch:16}.
\textchange{In fact, the proof of the dichotomy theorem by Han-Kobayashi \cite{HanKob87} crucially uses the CI structure.
More specifically, the Han-Kobayashi condition induces the CI structure for i.i.d. sources with positivity condition, which facilitate single-letter characterization.
Gel'fand-Pinsker \cite{GelPin:79} also used the CI structure to derive single-letter characterization of the multiterminal source coding problem
including the multiterminal distributed function computing problem as a special case.
Focusing on the CI structure in this paper is inspired by ideas in these prior works.
However, in our approach, the CI structure is used ``operationally" to decouple the coding system into multiple parts of coding systems with full side-information,
which facilitates using of the proof method based on informative structure.
As long as the author know, such an operational usage of the CI  structure in converse proof has never appeared in the literature,
and is of independent interest.}

Another new element is recursion. When we apply the above mentioned two ingredients to a given function, 
we can show that the function computation region of the given function is included in
the function computation region of a ``finer" function.\footnote{\textchange{A function $f(x)$ is finer than function $g(x)$ when there exists a function $h$ 
such that $g(x) = h(f(x))$.}} If the finer function is tantamount to the identity function, then
we can conclude that the function computation region of the original function coincides with the Slepian-Wolf region.
However, even when the finer function is not the identity function, we can apply the same argument to that finer function
to show that the function computation region of the finer function is included in the function computation region
of a further finer function. By repeating this procedure recursively, we can eventually show that
the function computation region of the original function coincides with the Slepian-Wolf region.

\textchange{As an application of our general approach, when the number of terminals is two, 
we reproduce the dichotomy theorem of Han-Kobayashi \cite{HanKob87} and that of Kuzuoka-Watanabe \cite{KuzWat15}
in a unified manner. In addition to reproducing these results, our approach can provide converse results for cases  
that cannot be covered by neither of the methods in \cite{HanKob87} nor \cite{KuzWat15},
though we do not pursue the two terminals problem in depth in this paper; we will provide one of such examples, a mixture of i.i.d. sources,
in Section \ref{subsec:mixture}.  
We also apply our approach to derive dichotomy theorems
for more than two terminals. In fact, for the the class of i.i.d. sources with
positivity condition, we derive a sufficient condition that strictly subsume the sufficient condition shown in \cite{HanKob87}. 
Furthermore, for the class of smooth sources, we establish the complete characterization of the classification problem. 
Note that the solution of classification problem for smooth sources does not imply the solution
of classification problem for i.i.d. sources with positivity condition since the latter problem is not a special case 
of the former problem. As we mentioned above, even when the optimal function computing region coincides with the SW region
for every i.i.d. sources with positivity condition, the SW region may be improved for sources having memory.}

\subsection*{Organization of Paper}

The rest of the paper is organized as follows:
In Section \ref{sec:problem-formulation}, we introduce the problem formulation of
multiterminal distributed computing. In Section \ref{section:overview}, we explain 
an overview of our approach, and in Section \ref{section:general-results}, we state and prove our general results. 
In Section \ref{section:two-terminal}, we revisit known results on two-terminal distributed computing from the view
point of our general approach. In Section \ref{section:multi-terminal-iid} and Section \ref{section:multi-terminal-smooth},
we apply our approach to the multiterminal distributed computing for the class of i.i.d. sources with positivity condition
and the class of smooth sources, respectively.
We close the paper with some conclusion in Section \ref{section:conclusion}.
Some of technical results and proofs are given in appendices.

\subsection*{Notations}

Throughout this paper, random variables (e.g., $X$) and their realizations (e.g., $x$) are denoted by capital 
and lower case letters, respectively. All random variables take values in some finite alphabets which are denoted by
the respective calligraphic letters (e.g., ${\cal X}$). The probability distribution of random variable $X$ is denoted by $P_X$.
Similarly, $X^n := (X_1,X_2,\ldots,X_n)$ and $x^n := (x_1,x_2,\ldots,x_n)$ denote, respectively, a random vector and its
realization in the $n$th Cartesian product ${\cal X}^n$ of ${\cal X}$. We will use bold lower letters to
represent vectors if the length $n$ is apparent from the context; e.g., we use $\bm{x}$ instead of $x^n$.

For a finite set ${\cal S}$, the cardinality of ${\cal S}$ is denoted by $|{\cal S}|$. 
For a subset ${\cal T} \subseteq {\cal S}$, the complement ${\cal S} \backslash {\cal T}$ is denoted by ${\cal T}^c$.
For a given sequence $\bm{s}$ in the $n$th Cartesian product ${\cal S}^n$
of ${\cal S}$, the type $P_{\bm{s}} = (P_{\bm{s}}(s) : s \in {\cal S})$ of $\bm{s}$ is defined by
\begin{align}
P_{\bm{s}}(s) := \frac{|\{ i \in [1:n] : s_i = s\}|}{n},~~~s \in {\cal S},
\end{align}
where $[1:n] := \{1,2,\ldots,n\}$. The set of all types of sequences in ${\cal S}^n$ is denoted by ${\cal P}_n({\cal S})$.
The indicator function is denoted by $\mathbf{1}[\cdot]$.
Information-theoretic quantities are denoted in the usual manner \cite{cover, csiszar-korner:11}. 
The binary entropy is denoted by $h(\cdot)$. All logarithms are with respect to base $2$. 
For given distributions $P, Q$, the variational distance is denoted by
$\| P - Q \|_1 := \frac{1}{2} \sum_a |P(a) - Q(a)|$.

\section{Problem Formulation} \label{sec:problem-formulation}

Let 
\begin{align}
(\bm{X}_1,\ldots,\bm{X}_L) = \{ (X_1^n,\cdots,X_L^n) \}_{n=1}^\infty
\end{align}
be a general correlated source with finite alphabet ${\cal X}_{{\cal L}} = {\cal X}_1 \times \cdots \times {\cal X}_L$; the source is
general in the sense of \cite{han:93} (see also \cite{han:book}), i.e., it may have memory and may not be stationary nor ergodic. Without loss of generality,
we assume ${\cal X}_\ell = \{ 0, 1, \ldots, |{\cal X}_\ell | -1 \}$ for $\ell =1,\ldots,L$. For a subset ${\cal A} \subseteq {\cal L}$
of the set ${\cal L} = \{1,\ldots, L\}$ of all terminals, we denote $\bm{X}_{\cal A} = (\bm{X}_\ell : \ell \in {\cal A})$,
$X_{\cal A}^n = (X_\ell^n : \ell \in {\cal A})$, and etc. 

We consider a sequence $\bm{f} = \{ f_n \}_{n=1}^\infty$ of functions $f_n : {\cal X}_{\cal L}^n \to {\cal Z}_n$. 
A code $\Phi_n = (\varphi_n^{(1)},\ldots,\varphi_n^{(L)}, \psi_n)$ for computing $f_n$ is defined by encoders 
$\varphi_n^{(\ell)}:{\cal X}_\ell^n \to {\cal M}_n^{(\ell)}$ for $\ell = 1,\ldots,L$ and a decoder $\psi_n: {\cal M}_n^{(1)}\times \cdots \times {\cal M}_n^{(L)} \to {\cal Z}_n$.
The error probability of the code $\Phi_n$ is given by
\begin{align}
\Pe(\Phi_n | f_n) := \Pr\left( \psi_n(\varphi_n^{(\ell)}(X_\ell^n) : \ell \in {\cal L}) \neq f_n(X_\ell^n:\ell \in {\cal L}) \right).
\end{align} 

\begin{definition}
For a given source $\bm{X}_{\cal L}$ and a sequence of functions $\bm{f}$, a rate tuple $R_{\cal L} = (R_\ell : \ell \in {\cal L})$ is defined to be achievable
if there exists a sequence $\{ \Phi_n \}_{n=1}^\infty$ of codes satisfying 
\begin{align}
\lim_{n\to\infty} \Pe(\Phi_n|f_n) = 0
\end{align}
and
\begin{align}
\limsup_{n\to\infty} \frac{1}{n} \log |{\cal M}_n^{(\ell)}| \le R_\ell,~~~\forall \ell \in {\cal L}.
\end{align}
The achievable rate region for computing $\bm{f}$, denoted by ${\cal R}(\bm{X}_{\cal L}|\bm{f})$, is the set of all achievable rate tuples. 
\end{definition}

One of the most simple classes of functions is the class of symbol-wise functions defined as follows;
we will investigate this class intensively later in Sections \ref{section:two-terminal}-\ref{section:multi-terminal-smooth}.

\begin{definition}[Symbol-wise Function]
For a given function $f:{\cal X}_{\cal L} \to {\cal V}$, a function defined as 
$f_n(\bm{x}_{\cal L}) = (f(x_{{\cal L},1}), \ldots,f(x_{{\cal L},n}))$ is called {\em symbol-wise} function
induced by $f$. We sometimes denote $f^n$ instead of $f_n$ to emphasize that the symbol-wise function
is $n$ copies of $f$.
\end{definition}

In order to compute functions, one approach is that all terminals send messages so 
that the decoder can reproduce the entire source, which is known as the Slepian-Wolf coding.

\begin{definition}[SW Region]
For a given source $\bm{X}_{\cal L}$, the achievable rate region ${\cal R}(\bm{X}_{\cal L} | \bm{f}^\san{id})$ for the sequence
$\bm{f}^\san{id} = \{ f_n^{\san{id}} \}_{n=1}^\infty$ of identity functions is called the Slepian-Wolf (SW) region, and denoted by ${\cal R}_{\san{SW}}(\bm{X}_{\cal L})$.
\end{definition}
Note that ${\cal R}_{\san{SW}}(\bm{X}_{\cal L})$ is a trivial inner bound on ${\cal R}(\bm{X}_{\cal L}|\bm{f})$, i.e., it holds that
${\cal R}_{\san{SW}}(\bm{X}_{\cal L}) \subseteq {\cal R}(\bm{X}_{\cal L}|\bm{f})$ for any $\bm{f}$.

The Slepian-Wolf coding for general sources was first studied in \cite{miyake:95} (see also \cite[Section 7.2]{han:book}).
Even though only the two terminal case was studied in the literature, it is straightforward to extend the result to the multiterminal case;
the SW region for general sources is characterized as follows.
\begin{proposition} \label{proposition:SW-region}
For a given general source $\bm{X}_{\cal L}$, it holds that 
\begin{align} \label{eq:SW-region-characterization}
{\cal R}_{\san{SW}}(\bm{X}_{\cal L}) = \bigg\{ R_{\cal L} : \sum_{\ell \in {\cal A}} R_\ell \ge \overline{H}(\bm{X}_{\cal A}|\bm{X}_{{\cal A}^c}),~\forall {\cal A} \subseteq {\cal L} \bigg\},
\end{align}
where 
\begin{align}
\overline{H}(\bm{X}_{{\cal A}}|\bm{X}_{{\cal A}^c}) := \inf\bigg\{ \alpha : \lim_{n\to\infty} \Pr\bigg\{ \frac{1}{n} \log \frac{1}{P_{X^n_{\cal A}|X^n_{{\cal A}^c}}(X^n_{\cal A}|X^n_{{\cal A}^c})} > \alpha \bigg\} = 0 \bigg\}
\end{align}
is the spectral conditional sup-entropy rate of $\bm{X}_{\cal A}$ given $\bm{X}_{{\cal A}^c}$.
\end{proposition}
When the source is i.i.d., the characterization in Proposition \ref{proposition:SW-region}
reduces to the well known formula by replacing $\overline{H}(\bm{X}_{\cal A}|\bm{X}_{{\cal A}^c})$ with the conditional entropy $H(X_{\cal A}|X_{{\cal A}^c})$
in \eqref{eq:SW-region-characterization}.

In this paper, we shall develop an approach to characterize the function computation region ${\cal R}(\bm{X}_{\cal L}|\bm{f})$ for general sources.
The function computation region can be sensitive to the support of distributions (cf.~the comment at the end of \cite[Sec. III]{HanKob87}).
To avoid such a complication, we consider the following class of smooth sources introduced in \cite{KuzWat15}.

\begin{definition}[Smooth Source]
A general source $\bm{X}_{\cal L}$ is said to be {\em smooth} if there exists a constant $0 < q < 1$, which does not depend on $n$, satisfying 
\begin{align}
P_{X^n_{\cal L}}(\bm{x}_{\cal L}) \ge q P_{X^n_{\cal L}}(\hat{\bm{x}}_{\cal L})
\end{align}
for every $\bm{x}_{\cal L},\hat{\bm{x}}_{\cal L} \in {\cal X}^n_{\cal L}$ with $d_H(\bm{x}_{\cal L}, \hat{\bm{x}}_{\cal L}) = 1$, where $d_H(\cdot,\cdot)$ is the Hamming distance. 
\end{definition}

The class of smooth sources is a natural generalization of i.i.d. sources with
positivity condition studied in \cite{HanKob87, AhlCsi:81}, and enables us to study distributed 
computation for a variety of sources in a unified manner. Indeed, this class contains 
sources with memory, such as Markov sources with positive transition matrices, or
non-ergodic sources, such as mixtures of i.i.d. sources with positivity condition;
see \cite{KuzWat15} for the detail.

\section{An Overview of The Approach} \label{section:overview}

\begin{table}[t]
\centering{
\caption{$f:{\cal X}_1\times {\cal X}_2 \to {\cal V}$}
\begin{tabular}{c|c|c|c|} \label{table:Example-overview}
 $x_1 \setminus x_2$ & 0 & 1& 2  \\\hline
 0 & 0 & 3 & 3  \\ \hline
 1 & 0 & 4 & 2  \\ \hline
 2 & 1 & 1 & 2  \\ \hline
\end{tabular}
}
\end{table}

In this section, we shall explain an overview of our approach for the special 
case of the two terminals and symbol-wise functions.
In our approach, we use three ingredients: the proof method based on informative structure \cite{KuzWat16}, the conditional independence structure, and the recursion. 
Since the third one, recursion, is not needed in the two terminals case, we only explain the first two ingredients here. 
For concreteness, let us consider symbol-wise function given by Table \ref{table:Example-overview}.
\textchange{Since this function satisfies the Han-Kobayashi condition \cite{HanKob87} (see also Section \ref{section:two-terminal}), the function computation region ${\cal R}(\bm{X}_{\cal L}|\bm{f})$
for this symbol-wise function coincides with the Slepian-Wolf region ${\cal R}_{\san{SW}}(\bm{X}_{\cal L})$ for any i.i.d. sources with positivity condition.
As an overview of our approach, we shall provide an alternative proof of this known fact. }

\begin{figure}[t]
\centering{
\includegraphics[width=0.45\textwidth, bb=0 0 283 85]{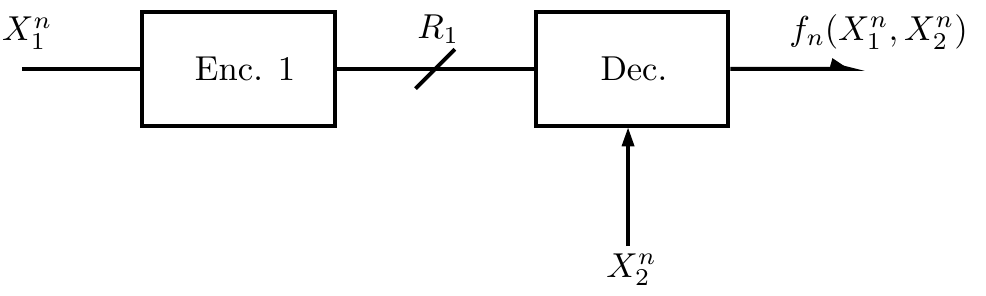}
\caption{A description of distributed computing with full side-information.}
\label{Fig:system-two-full-side}
}
\end{figure}

Let us first consider the special case where $X_2^n$ is observed at the decoder as full side-information
(see Fig.~\ref{Fig:system-two-full-side}). For this special case, the following approach to prove $R_1 \ge H(X_1|X_2)$
was proposed in \cite{KuzWat16}. Let $\Phi_n = (\varphi_n^{(1)},\psi_n)$ be a code to compute $f_n$ with vanishing error probability $\varepsilon_n$.
Since message $\varphi_n^{(1)}(X^n_1)$ is encoded without knowing the realization of side-information $X_2^n$, if we input $\pi_i^b(X_2^n)$ to
$\psi_n(\varphi_n^{(1)}(X^n_1), \cdot)$ instead of $X_2^n$, then we expect it will outputs
$f_n(X_1^n, \pi_i^b(X_2^n))$ with high probability, where $\pi_i^b$ shifts $i$th symbol of $X_2^n$ by $b$, i.e., $X_{2,i} \mapsto X_{2,i} + b \pmod{|{\cal X}_2|}$.
In fact, by noting the smoothness of the source, i.e., the fact that the probability does not change drastically when one symbol is shifted, we can show 
\begin{align}
\Pr\bigg( \psi_n(\varphi_n^{(1)}(X^n_1), \pi_i^b(X_2^n)) \neq  f_n(X_1^n, \pi_i^b(X_2^n)) \bigg) \le \frac{\varepsilon_n}{q}.
\end{align}
This implies that, via union bound, the decoder can reproduce the list $\big(  f(X_{1,i}, x_2) : x_2 \in {\cal X}_2 \big)$
with vanishing error probability. Here, note that every rows in Table \ref{table:Example-overview} are different. Thus, the decoder can
distinguish $X_{1,i}$ from the list. By doing this procedure for $1 \le i \le n$, the decoder can reproduce the entire source $X_1^n$
with vanishing error probability.\footnote{For simplicity, we have omitted one step in which we boost the small symbol error probability
to the small block error probability (see Lemma \ref{lemma:boosting}).} This means that the rate $R_1$ must be as large as the Slepian-Wolf rate $H(X_1|X_2)$.
A similar argument can be applied to prove $R_2 \ge H(X_2|X_1)$ since every columns in Table \ref{table:Example-overview} are different.
This is the first ingredient, informative structure, i.e., the function has a structure that reveals information about $X_{\ell,i}$.

\begin{figure}[t]
\centering{
\includegraphics[width=0.45\textwidth, bb=0 0 283 85]{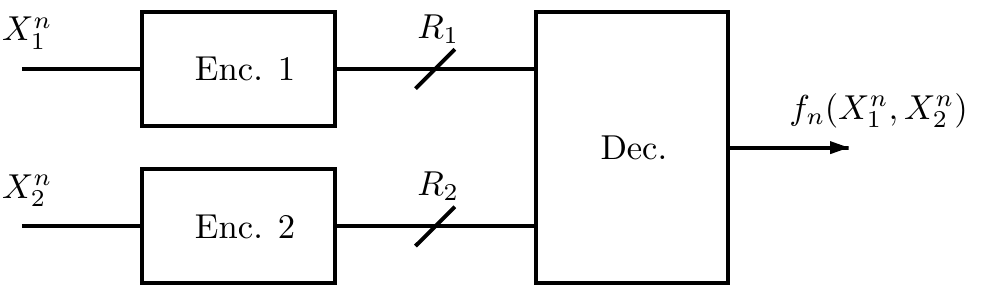}
\caption{A description of two-terminal distributed computing.}
\label{Fig:system-two-region}
}
\end{figure}

\begin{figure}[t]
\centering{
\includegraphics[width=0.45\textwidth, bb=0 0 283 165]{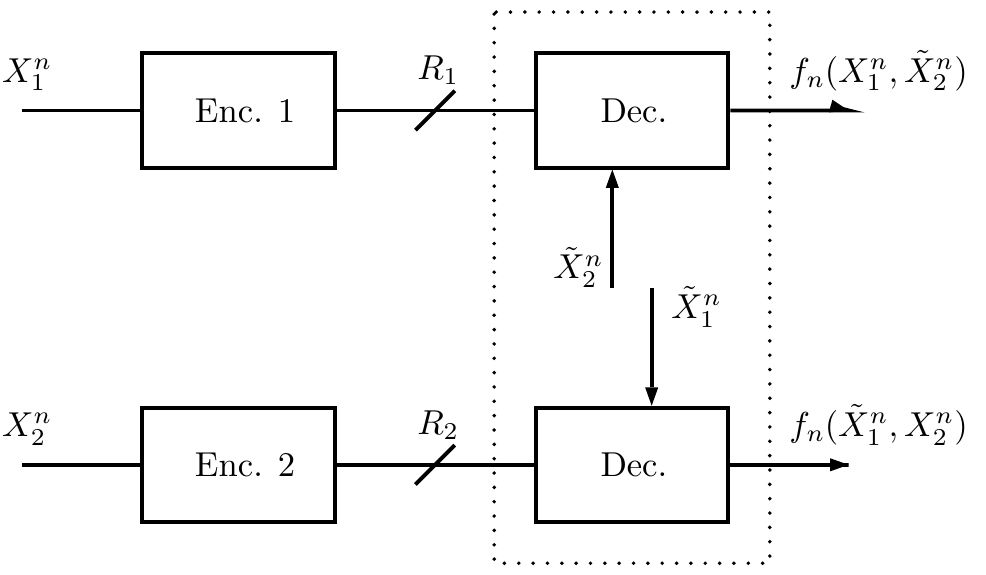}
\caption{A description of two virtual systems decoupled from the system in Fig.~\ref{Fig:system-two-region}.}
\label{Fig:system-decouple}
}
\end{figure}

In order to consider the entire region ${\cal R}(\bm{X}_{\cal L}|\bm{f})$, we need the second ingredient, conditional independence structure.
In the above argument to bound individual rates, a crucial step is to input $\pi_i^b(X_2^n)$ to $\psi_n(\varphi_n^{(1)}(X^n_1), \cdot)$;
this is possible because $X_2^n$ is directly observed at the decoder in the system of Fig.~\ref{Fig:system-two-full-side}.
When we consider the entire region, the same argument cannot be used since the decoder only gets $\varphi_n^{(1)}(X_1^n)$
and $\varphi_n^{(2)}(X_2^n)$.
In order to circumvent this problem, we note that the function in Table \ref{table:Example-overview} induces conditional independence, i.e., the Markov chain
$X_1 \markov f(X_1,X_2) \markov X_2$ holds.\footnote{\textchange{Note that this CI structure is also used in the proof of  \cite{HanKob87} to facilitate single letter characterization.}} 
In fact, it can be verified from the fact that either $x_1$ or $x_2$ is determined from $f(x_1,x_2)$.
Because of this observation, once the decoder can reproduce $V^n = f_n(X_1^n,X_2^n)$ (with vanishing error probability), 
the decoder can generate $\tilde{X}_1^n$ and $\tilde{X}_2^n$ via channels $P_{X_1|V}^n$ and $P_{X_2|V}^n$ so that 
$(X_1^n,\tilde{X}_2^n)$ and $(\tilde{X}_1^n, X_2^n)$ have (approximately) the same joint distribution as $(X_1^n,X_2^n)$.
By this argument, we can decouple the entire system (see Fig.~\ref{Fig:system-two-region}) into two virtual systems (see Fig.~\ref{Fig:system-decouple}).
Then, we can use the previous argument of the full side-information case, and show that the decoder can reproduce 
the entire sources $X_1^n$ and $X_2^n$ with vanishing error probability. This means that the function computation region ${\cal R}(\bm{X}_{\cal L}|\bm{f})$
must be included in the Slepian-Wolf region ${\cal R}_{\san{SW}}(\bm{X}_{\cal L})$.

In the above argument, the function being symbol-wise and the source being i.i.d. are
not necessary. In fact, the argument goes thorough as long as the function reveals (partial) information
about $X_{\ell,i}$  and the function induces conditional independence for the given source.
In the next section, we will introduce an approach to derive converse bounds on the multiterminal distributed computing 
by generalizing the above argument.

\section{General Result} \label{section:general-results}

For $\ell \in {\cal L}$, let $\overline{{\cal X}}_\ell$ be a partition of ${\cal X}_\ell$; i.e., $\overline{{\cal X}}_\ell = \{ {\cal C}_1,\ldots,{\cal C}_t\}$ is a set of nonempty subsets 
${\cal C}_i \subseteq {\cal X}_\ell~(i=1,\ldots,t)$ satisfying ${\cal C}_i \cap {\cal C}_j = \emptyset~(i\neq j)$ and
${\cal X}_\ell = \cup_{{\cal C} \in \overline{{\cal X}}_\ell} {\cal C}$. 
We say that partition $\overline{{\cal X}}_\ell$ is nontrivial if $t \ge 2$.
For each $x_\ell \in {\cal X}_\ell$, the subset ${\cal C} \in \overline{{\cal X}}_\ell$ satisfying
$x_\ell \in {\cal C}$ is uniquely determined and denoted by $[x_\ell]_{\overline{{\cal X}}_\ell}$. For a sequence $\bm{x}_\ell = (x_{\ell,1},\ldots,x_{\ell,n}) \in {\cal X}_\ell^n$,
we apply $[\cdot]_{\overline{{\cal X}}_\ell}$ to each symbol, i.e.,
$[\bm{x}_\ell]_{\overline{{\cal X}}_\ell} = ([x_{\ell,1}]_{\overline{{\cal X}}_\ell},\ldots,[x_{\ell,n}]_{\overline{{\cal X}}_\ell}) \in \overline{{\cal X}}_\ell^n$. 
For a subset ${\cal A} \subseteq {\cal L}$ and a tuple $\overline{{\cal X}}_{\cal A} = (\overline{{\cal X}}_\ell : \ell \in {\cal A})$
of partitions, we denote $[x_{{\cal A}}]_{\overline{{\cal X}}_{\cal A}} = ([x_\ell]_{\overline{{\cal X}}_\ell} : \ell \in {\cal A})$
and $[\bm{x}_{\cal A}]_{\overline{{\cal X}}_{\cal A}} = ([\bm{x}_\ell]_{\overline{{\cal X}}_\ell} : \ell \in {\cal A})$.

For a subset ${\cal A} \subseteq {\cal L}$, 
a symbol $a_{\cal A} \in {\cal X}_{\cal A}$, a sequence $\bm{x}_{\cal A} \in {\cal X}_{\cal A}^n$, and an index $i \in [1:n]$, 
let $a_{\cal A} \bm{x}_{\cal A}^{(-i)}$ be the sequence such that elements $x_{{\cal A},i}$ of $\bm{x}_{\cal A}$ are replaced by $a_{\cal A}$.

In the following, we introduce a few structures of functions; we say that a sequence $\bm{f} = \{ f_n \}_{n=1}^\infty$ of functions has
certain structure if $f_n$ has that structure for every $n$.

\begin{definition}[Local Function]
For a given tuple of partitions $\overline{{\cal X}}_{\cal L} = ( \overline{{\cal X}}_\ell : \ell \in {\cal L})$, 
the symbol-wise function $f_{n,\overline{{\cal X}}_{\cal L}} : {\cal X}_{{\cal L}}^n \to \prod_{\ell=1}^L \overline{{\cal X}}_\ell^n$
defined by
\begin{align} \label{eq:local-function}
f_{n,\overline{{\cal X}}_{\cal L}}(\bm{x}_\ell : \ell \in {\cal L}) := ( [\bm{x}_\ell]_{\overline{{\cal X}}_\ell} : \ell \in {\cal L} ).
\end{align}
is called {\em local function}.
\end{definition}

Since a local function is a symbol-wise function, we sometimes just say local function $f_{\overline{{\cal X}}_{\cal L}}:{\cal X}_{{\cal L}} \to \prod_{\ell=1}^L \overline{{\cal X}}_\ell$.
As is clear from the definition, $\ell$th component of local functions 
can be locally computable at $\ell$th terminal. 
For a sequence of local function $\bm{f}_{\overline{{\cal X}}_{\cal L}} = \{ f_{n, \overline{{\cal X}}_{\cal L}} \}_{n=1}^\infty$, the achievable rate region
${\cal R}(\bm{X}_{\cal L} | \bm{f}_{\overline{{\cal X}}_{\cal L}})$ plays an important role. Note that ${\cal R}(\bm{X}_{\cal L} | \bm{f}_{\overline{{\cal X}}_{\cal L}}) = {\cal R}_{\san{SW}}(\bm{X}_{\cal L})$
when all the partitions are the finest partitions, i.e., $\overline{{\cal X}}_\ell = \{\{ x_\ell \} : x_\ell \in {\cal X}_\ell\}$ for every $\ell \in {\cal L}$. In the rest of the paper, we write
$\overline{{\cal X}}_\ell \equiv {\cal X}_\ell$ when $\overline{{\cal X}}_\ell$ is the finest partition. 

The following class of functions is a multiterminal generalization of the function class introduced in \cite{KuzWat16};
for the motivation of the definition, see \cite[Sections III and IV]{KuzWat16}.

\begin{definition}[Informative Function] \label{definition:informative-function}
For a subset ${\cal A} \subseteq {\cal L}$, let $\overline{{\cal X}}_{\cal A} = (\overline{{\cal X}}_\ell : \ell \in {\cal A})$ be a tuple of partitons. A function $f_n$
is said to be $\overline{{\cal X}}_{\cal A}$-informative if $f_n$ satisfies the following conditions:
\begin{enumerate}
\item \label{condition:informative-1}
For each $i \in [1:n]$, there exists a mapping $\xi^{(i)}_{n,{\cal A}}: {\cal Z}_n^{|{\cal X}_{{\cal A}^c}|} \to \overline{{\cal X}}_{\cal A}$ such that,\footnote{By a slight abuse of notation,
we also use notation $\overline{{\cal X}}_{\cal A}$ to describe $\times_{\ell \in {\cal A}} \overline{{\cal X}}_\ell$ though the same notation is used for a tuple of partitions.} 
for any $a_{\cal A} = (a_\ell : \ell \in {\cal A}) \in {\cal X}_{\cal A}$ and
$\bm{x}_{\cal L} \in {\cal X}_{\cal L}^n$, 
\begin{align} \label{eq:condition-informative-1}
\xi^{(i)}_{n,{\cal A}}\left( \left( f_n( a_{\cal A} \bm{x}_{\cal A}^{(-i)}, a_{{\cal A}^c} \bm{x}_{{\cal A}^c}^{(-i)}) : a_{{\cal A}^c} \in {\cal X}_{{\cal A}^c} \right) \right) 
 =  [ a_{\cal A} ]_{\overline{{\cal X}}_{\cal A}}  
 =  ([a_\ell]_{\overline{{\cal X}}_\ell} : \ell \in {\cal A}).
\end{align}

\item For every $\bm{x}_{\cal L} \in {\cal X}_{\cal L}^n$, every $\ell \in {\cal A}$, and any permutation $\sigma$ on $[1:n]$ satisfying 
$[\sigma(\bm{x}_\ell)]_{\overline{{\cal X}}_\ell} = [\bm{x}_\ell]_{\overline{{\cal X}}_\ell}$, 
\begin{align}
f_n(\bm{x}_1,\ldots,\sigma(\bm{x}_\ell),\ldots,\bm{x}_L) = f_n(\bm{x}_1,\ldots,\bm{x}_\ell,\ldots,\bm{x}_L).
\end{align}
\end{enumerate}
On the other hand, when only Condition \ref{condition:informative-1} is satisfied, $f_n$ is said to be {\em semi $\overline{{\cal X}}_{\cal A}$-informative}.
\end{definition}

\begin{remark}
In contrast to the two terminal case in \cite{KuzWat16}, when we say ``$\overline{{\cal X}}_{\cal A}$-informative",
it also designates the coordinates ${\cal A}^c$ we can vary in \eqref{eq:condition-informative-1}.
\end{remark}

\begin{remark}
Condition (\ref{condition:informative-1}) means that, for each $i \in [1:n]$ and for any $a_{\cal A} = (a_\ell : \ell \in {\cal A})  \in {\cal X}_{\cal A}$,
the tuple of equivalence classes $({\cal K}_\ell : \ell \in {\cal A})$ satisfying $a_\ell \in {\cal K}_\ell$ for each $\ell \in {\cal A}$
can be uniquely determined from the list 
\begin{align}
\left( f_n( a_{\cal A} \bm{x}_{\cal A}^{(-i)}, a_{{\cal A}^c} \bm{x}_{{\cal A}^c}^{(-i)}) : a_{{\cal A}^c} \in {\cal X}_{{\cal A}^c} \right)
\end{align}
irrespective of $\bm{x}_{\cal L} \in {\cal X}_{\cal L}^n$.
From the definition, it is apparent that, if a function is $\overline{{\cal X}}_{\cal A}$-informative, 
then it is $\overline{{\cal X}}_{{\cal A}^\prime}$-informative for every ${\cal A}^\prime \subseteq {\cal A}$.
\end{remark}

For two terminals setting, three examples of informative functions are given in \cite[Section IV]{KuzWat16};
the first one is symbol-wise functions, and the latter two are not necessarily symbol-wise functions. 
For later convenience, we would like to review the first one here; the multiterminal generalization will be studied 
in Sections \ref{section:multi-terminal-iid} and \ref{section:multi-terminal-smooth}.

For a given function $f:{\cal X}_1 \times {\cal X}_2 \to {\cal V}$, let $\overline{{\cal X}}_{1,f}$ be the partition of ${\cal X}_1$
such that two symbols $x_1$ and $\hat{x}_1$ are in the same partition of ${\cal X}_1$ if and only if
$f(x_1,x_2) = f(\hat{x}_1,x_2)$ for every $x_2 \in {\cal X}_2$. Similarly, we define partition $\overline{{\cal X}}_{2,f}$
of ${\cal X}_2$ by interchanging the role of the two terminals. 
\textchange{It is known that the equivalence class given by the partition $\overline{{\cal X}}_{i,f}$ plays an important role
to characterize the optimal rate for computing the function; it represents the equivalence class of the encoder's observation
that is learnable by the decoder when the function is computed \cite{OrlRoc:01, KowKum12}.}
For the function given by Table \ref{table:Example-2terminal},
the partitions are given by $\overline{{\cal X}}_{1,f} = \{ \{0\}, \{1,2\}\}$
and $\overline{{\cal X}}_{2,f} \equiv {\cal X}_2$. 
For two terminal symbol-wise functions, we have the following characterization.
\begin{proposition}[\cite{KuzWat16}] \label{proposition:informative-symbol-wise}
A symbol-wise function $f_n:{\cal X}_1^n \times {\cal X}_2^n \to {\cal V}^n$
defined from $f:{\cal X}_1 \times {\cal X}_2 \to {\cal V}$ is $\overline{{\cal X}}_{1,f}$-informative
and $\overline{{\cal X}}_{2,f}$-informative, respectively.
\end{proposition}

\begin{table}[tb]
\centering{
\caption{$f:{\cal X}_1\times {\cal X}_2 \to {\cal V}$}
\begin{tabular}{c|c|c|c|} \label{table:Example-2terminal}
 $x_1 \setminus x_2$ & 0 & 1& 2  \\\hline
 0 & 0 & 1 & 2  \\ \hline
 1 & 3 & 4 & 2  \\ \hline
 2 & 3 & 4 & 2  \\ \hline
\end{tabular}
}
\end{table}

Next, we shall introduce the class of functions that induce conditional independence.
\begin{definition}[Conditional Independence]
For a given source $\bm{X}_{\cal L}$  and a nontrivial partition $\overline{{\cal L}}$ of ${\cal L}$, 
we say that $\bm{f} = \{f_n \}_{n=1}^\infty$ induces (approximate) conditional independence for $(\bm{X}_{\cal L}, \overline{{\cal L}})$ if there exists a sequence
of random variables $\bm{S} = \{ S_n \}_{n=1}^\infty$ such that, for every $n \ge 1$, 
\begin{align}
X_{{\cal A}}^n \markov S_n \markov X_{{\cal A}^c}^n
\end{align}
for every ${\cal A} \in \overline{{\cal L}}$ and
\begin{align} \label{eq:conditional-independence-2}
\Pr( \gamma_n(Z_n) \neq S_n) \le \mu_n
\end{align}
for some function $\gamma_n:{\cal Z}_n \to {\cal S}_n$ with $\mu_n \to 0$ as $n\to \infty$. 
\end{definition}

Note that whether conditional independence is induced or not 
depend on both functions and sources in general. When the source $\bm{X}_{\cal L}$ is independent across a partition
$\overline{{\cal L}}$, then any function induces conditional independence for $(\bm{X}_{\cal L}, \overline{{\cal L}})$
by taking $\bm{S}$ to be a constant. In latter sections, we will show some conditions of functions
such that those functions induce conditional independence for $(\bm{X}_{\cal L}, \overline{{\cal L}})$
for every sources in the class of i.i.d. sources or the class of smooth sources. 

Finally, we shall introduce product functions. 
\begin{definition}[Product Function]
For given functions $\bm{f}_1 = \{f_{1,n}\}_{n=1}^\infty$ and $\bm{f}_2 = \{f_{2,n} \}_{n=1}^\infty$,
let us define the product function as follows:
\begin{align}
(f_{1,n},f_{2,n})(\bm{x}_{\cal L}) := (f_{1,n}(\bm{x}_{\cal L}), f_{2,n}(\bm{x}_{\cal L})).
\end{align}
For the product function $(\bm{f}_1,\bm{f}_2) = \{ (f_{1,n},f_{2,n}) \}_{n=1}^\infty$, the achievable rate
region is denoted by ${\cal R}(\bm{X}_{\cal L}|\bm{f}_1,\bm{f}_2)$.
The product of multiple functions is defined similarly.
\end{definition}

Now, we are ready to state our main results.

\begin{theorem} \label{theorem:general}
For a given smooth source $\bm{X}_{\cal L}$ and a nontrivial partition $\overline{{\cal L}}$ of ${\cal L}$, suppose that 
$\bm{f} = \{ f_n \}_{n=1}^\infty$ induces conditional independence for $(\bm{X}_{\cal L}, \overline{{\cal L}})$.
Furthermore, for a given tuple $\overline{{\cal X}}_{\cal L} = (\overline{{\cal X}}_\ell : \ell \in {\cal L})$ of partitions, suppose that 
$\bm{f}$ is $\overline{{\cal X}}_{\cal A}$-informative for every ${\cal A} \in \overline{{\cal L}}$, where $\overline{{\cal X}}_{\cal A} = (\overline{{\cal X}}_\ell : \ell \in {\cal A})$.
Then, we have
\begin{align} \label{theorem:main-informative}
{\cal R}(\bm{X}_{\cal L} | \bm{f}) = {\cal R}(\bm{X}_{\cal L} | \bm{f}_{\overline{{\cal X}}_{\cal L}}),
\end{align}
where $\bm{f}_{\overline{{\cal X}}_{\cal L}}$ is the local function defined by \eqref{eq:local-function} for the tuple $\overline{{\cal X}}_{\cal L}$.
On the other hand, if $\bm{f}$ is semi $\overline{{\cal X}}_{\cal A}$-informative for every ${\cal A} \in \overline{{\cal L}}$, then we have
\begin{align} \label{theorem:main-semi}
{\cal R}(\bm{X}_{\cal L} | \bm{f}) = {\cal R}(\bm{X}_{\cal L} | \bm{f}, \bm{f}_{\overline{{\cal X}}_{\cal L}}).
\end{align}
\end{theorem}

We can apply the latter claim \eqref{theorem:main-semi} of Theorem \ref{theorem:general} 
recursively to obtain the following corollaries. 

\begin{corollary}[Recursion] \label{corollary:recursion}
Let $\bm{X}_{\cal L}$ be a smooth source. For a given sequence of nontrivial partitions $\{ \overline{{\cal L}}^{(i)} \}_{i=1}^k$ of ${\cal L}$ and
a given sequence of tuples of partitions $\{ \overline{{\cal X}}_{\cal L}^{(i)} = (\overline{{\cal X}}_\ell^{(i)} : \ell \in {\cal L}) \}_{i=1}^k$,
suppose that, for every $i=1,\ldots,k$, $(\bm{f}, \bm{f}_{\overline{{\cal X}}_{\cal L}^{(1)}},\ldots, \bm{f}_{\overline{{\cal X}}_{\cal L}^{(i-1)}})$
induces\footnote{For $i=1$, $(\bm{f}, \bm{f}_{\overline{{\cal X}}_{\cal L}^{(1)}},\ldots, \bm{f}_{\overline{{\cal X}}_{\cal L}^{(i-1)}})$ means $\bm{f}$.}
conditional independence for $(\bm{X}_{\cal L}, \overline{{\cal L}}^{(i)})$ and 
semi $\overline{{\cal X}}_{\cal A}^{(i)}$-informative for every ${\cal A} \in \overline{{\cal L}}^{(i)}$. Then, we have
\begin{align}
{\cal R}(\bm{X}_{\cal L} | \bm{f}) = {\cal R}(\bm{X}_{\cal L} | \bm{f},  \bm{f}_{\overline{{\cal X}}_{\cal L}^{(1)}},\ldots, \bm{f}_{\overline{{\cal X}}_{\cal L}^{(k)}}).
\end{align}
\end{corollary}

\begin{corollary} \label{corollary:recursion-SW}
Under the same condition as Corollary \ref{corollary:recursion}, if $\overline{{\cal X}}_\ell^{(k)} \equiv {\cal X}_\ell$ for every $\ell \in {\cal L}$, 
then we have 
\begin{align}
{\cal R}(\bm{X}_{\cal L} | \bm{f}) = {\cal R}_{\san{SW}}(\bm{X}_{\cal L}).
\end{align}
\end{corollary}

\subsection{Proof of Converse Part of Theorem \ref{theorem:general}}

We first prove $\subseteq$ part of \eqref{theorem:main-informative}.
Let $\{ \Phi_n = (\varphi_n^{(1)},\ldots,\varphi_n^{(L)},\psi_n) \}_{n=1}^\infty$ be a code for computing $\bm{f}$, and suppose that
it achieves a rate tuple $R_{\cal L} = (R_\ell : \ell \in {\cal L}) \in {\cal R}(\bm{X}_{\cal L}|\bm{f})$.
From this code, we shall construct a modified code for computing local function $\bm{f}_{\overline{{\cal X}}_{\cal L}}$.
By letting $Z_n = f_n(X_\ell^n : \ell \in {\cal L})$ and $\hat{Z}_n = \psi_n(\varphi_n^{(\ell)}(X_\ell^n) : \ell \in {\cal L})$, we have
\begin{align} \label{eq:proof-main-theorem-error-probability}
\Pr(Z_n \neq \hat{Z}_n) \le \varepsilon_n
\end{align} 
for some $\varepsilon_n \to 0$ as $n\to \infty$. Since $\bm{f}$ induces conditional independence for $(\bm{X}_{\cal L}, \overline{{\cal L}})$, we can find 
$\bm{S} = \{ S_n \}_{n=1}^\infty$ such that 
\begin{align}
P_{X_{\cal L}^n S_n} = P_{S_n} \prod_{{\cal A} \in \overline{{\cal L}}} P_{X_{\cal A}^n|S_n}.
\end{align}
Also, there exists a mapping $\gamma_n$ satisfying \eqref{eq:conditional-independence-2}.
Furthermore, since $\bm{f}$ is $\overline{{\cal X}}_{\cal A}$-informative for every ${\cal A} \in \overline{{\cal L}}$,
there exists function $\xi^{(i)}_{n,{\cal A}}: {\cal Z}_n^{|{\cal X}_{{\cal A}^c}|} \to \overline{{\cal X}}_{\cal A}$
satisfying \eqref{eq:condition-informative-1} for each ${\cal A} \in \overline{{\cal L}}$ and $i \in [1:n]$.

Upon receiving messages $m_{\cal L} = (m_\ell : \ell \in {\cal L})$ from the encoders, the modified decoder first computes $\hat{s}_n = \gamma_n(\psi_n(m_{\cal L}))$,
and, for each ${\cal A} \in \overline{{\cal L}}$, locally generate $\tilde{X}_{{\cal A}}^n \sim P_{X_{{\cal A}}^n|S_n}(\cdot| \hat{s}_n)$. 
Then, for each ${\cal A} \in \overline{{\cal L}}$,
the decoder computes
\begin{align}
\xi^{(i)}_{n,{\cal A}}\left( \left( \psi_n\big( m_{\cal A} , \varphi_n^{({\cal A}^c)}( a_{{\cal A}^c} \tilde{X}_{{\cal A}^c}^{(-i)}) \big) : a_{{\cal A}^c} \in {\cal X}_{{\cal A}^c} \right) \right), 
\end{align}
where $m_{\cal A} = (m_\ell : \ell \in {\cal A})$, $\varphi_n^{({\cal A}^c)}( a_{{\cal A}^c} \tilde{X}_{{\cal A}^c}^{(-i)}) = ( \varphi_n^{(\ell)}(a_\ell \tilde{X}_\ell^{(-i)}) : \ell \in {\cal A}^c)$,
and $a_\ell \tilde{X}_\ell^{(-i)}$ is the sequence obtained by replacing $i$th coordinate $\tilde{X}_{\ell,i}$ of $\tilde{X}_\ell^n$ with $a_\ell$.
We shall show that, for each ${\cal A} \in \overline{{\cal L}}$,  
\begin{align}
W_{{\cal A},i} = \xi^{(i)}_{n,{\cal A}}\left( \left( \psi_n\big( \varphi_n^{({\cal A})}(X_{\cal A}^n), \varphi_n^{({\cal A}^c)}( a_{{\cal A}^c} \tilde{X}_{{\cal A}^c}^{(-i)}) \big) : a_{{\cal A}^c} \in {\cal X}_{{\cal A}^c} \right) \right)
\end{align}
coincides with $[X_{{\cal A},i}]_{\overline{{\cal X}}_{\cal A}}$ with high probability. 

By applying Lemma \ref{lemma:property-of-variational-distance}, \eqref{eq:proof-main-theorem-error-probability} and \eqref{eq:conditional-independence-2} imply
\begin{align}
\| P_{X_{\cal A}^n S_n} - P_{X_{\cal A}^n \hat{S}_n} \|_1 
&\le \Pr( S_n \neq \hat{S}_n) \\
&\le \Pr( Z_n \neq \hat{Z}_n \mbox{ or } \gamma_n(Z_n) \neq S_n) \\
&\le \varepsilon_n + \mu_n,
\end{align}
where $\hat{S}_n = \gamma_n(\hat{Z}_n)$. Then, since taking marginal does not increase the variational distance, we have
\begin{align}
\| P_{X_{\cal A}^n X_{{\cal A}^c}^n} - P_{X_{\cal A}^n \tilde{X}_{{\cal A}^c}^n} \|_1
&\le \| P_{X_{\cal A}^n X_{{\cal A}^c}^n S_n }  - P_{X_{\cal A}^n \tilde{X}_{{\cal A}^c}^n \hat{S}_n } \|_1 \\
&= \left\| P_{X_{\cal A}^n S_n } \prod_{{\cal B} \in \overline{{\cal L}}\backslash \{ {\cal A} \}} P_{X_{\cal B}^n|S_n} 
   - P_{X_{\cal A}^n \hat{S}_n } \prod_{{\cal B} \in \overline{{\cal L}}\backslash \{ {\cal A} \}} P_{X_{\cal B}^n|S_n}   \right\|_1 \\
 &= \| P_{X_{\cal A}^n S_n} - P_{X_{\cal A}^n \hat{S}_n} \|_1 \\
 &\le \varepsilon_n + \mu_n. 
 \label{eq:variational-distance-bound}
\end{align}
By noting \eqref{eq:condition-informative-1}, if 
\begin{align}
\left( \psi_n\big( \varphi_n^{({\cal A})}(X_{\cal A}^n), \varphi_n^{({\cal A}^c)}( a_{{\cal A}^c} \tilde{X}_{{\cal A}^c}^{(-i)}) \big) : a_{{\cal A}^c} \in {\cal X}_{{\cal A}^c} \right) 
\end{align}
coincides with
\begin{align}
\left( f_n(X_{\cal A}^n, a_{{\cal A}^c} \tilde{X}_{{\cal A}^c}^{(-i)}) : a_{{\cal A}^c} \in {\cal X}_{{\cal A}^c} \right),
\end{align}
then $W_{{\cal A},i}$ coincides with $[X_{{\cal A},i}]_{\overline{{\cal X}}_{\cal A}}$. Thus, we have
\begin{align}
\lefteqn{ \Pr( W_{{\cal A},i} \neq  [X_{{\cal A},i}]_{\overline{{\cal X}}_{\cal A}}) } \\
&\le 
\Pr\left(  \left( \psi_n\big( \varphi_n^{({\cal A})}(X_{\cal A}^n), \varphi_n^{({\cal A}^c)}( a_{{\cal A}^c} \tilde{X}_{{\cal A}^c}^{(-i)}) \big) : a_{{\cal A}^c} \in {\cal X}_{{\cal A}^c} \right) 
 \neq \left( f_n(X_{\cal A}^n, a_{{\cal A}^c} \tilde{X}_{{\cal A}^c}^{(-i)}) : a_{{\cal A}^c} \in {\cal X}_{{\cal A}^c} \right) \right) \\
 &\le \Pr\left(  \left( \psi_n\big( \varphi_n^{({\cal A})}(X_{\cal A}^n), \varphi_n^{({\cal A}^c)}( a_{{\cal A}^c} X_{{\cal A}^c}^{(-i)}) \big) : a_{{\cal A}^c} \in {\cal X}_{{\cal A}^c} \right) 
 \neq \left( f_n(X_{\cal A}^n, a_{{\cal A}^c} X_{{\cal A}^c}^{(-i)}) : a_{{\cal A}^c} \in {\cal X}_{{\cal A}^c} \right) \right) + \varepsilon_n + \mu_n,
\end{align}
where the second inequality follows from Lemma \ref{lemma:probability-variational-distance} and \eqref{eq:variational-distance-bound}.
Furthermore, by using the smoothness of the source $X_{{\cal L}}^n$, we can show that (see Appendix \ref{appendix:smooth-bound})
\begin{align}
\Pr\left(  \left( \psi_n\big( \varphi_n^{({\cal A})}(X_{\cal A}^n), \varphi_n^{({\cal A}^c)}( a_{{\cal A}^c} X_{{\cal A}^c}^{(-i)}) \big) : a_{{\cal A}^c} \in {\cal X}_{{\cal A}^c} \right) 
 \neq \left( f_n(X_{\cal A}^n, a_{{\cal A}^c} X_{{\cal A}^c}^{(-i)}) : a_{{\cal A}^c} \in {\cal X}_{{\cal A}^c} \right) \right)
 \le \frac{|{\cal X}_{{\cal A}^c}|}{q} \varepsilon_n.
 \label{eq:smooth-bound}
\end{align}
Thus, we have
\begin{align}
\Pr( W_{{\cal A},i} \neq  [X_{{\cal A},i}]_{\overline{{\cal X}}_{\cal A}})
 \le \left( 1 + \frac{|{\cal X}_{{\cal A}^c}|}{q} \right) \varepsilon_n + \mu_n,
\end{align}
which implies 
\begin{align}
\Pr( W_{\ell,i} \neq  [X_{\ell,i}]_{\overline{{\cal X}}_\ell})
 \le \left( 1 + \frac{|{\cal X}_{{\cal A}^c}|}{q} \right) \varepsilon_n + \mu_n
\end{align}
for each $\ell \in {\cal A}$.
By conducting the above procedures for $i \in [1:n]$, $W_\ell^n = (W_{\ell,1},\ldots,W_{\ell,n})$ satisfies
\begin{align}
\mathbb{E}\left[ \frac{1}{n} d_H(W_{\ell}^n, [X_{\ell}^n]_{\overline{{\cal X}}_\ell})  \right]
&= \frac{1}{n} \sum_{i=1}^n \Pr( W_{\ell,i} \neq  [X_{\ell,i}]_{\overline{{\cal X}}_\ell}) \\
&\le \left( 1 + \frac{|{\cal X}_{{\cal A}^c}|}{q} \right) \varepsilon_n + \mu_n.
\end{align}
By the Markov inequality, for any $\beta > 0$, we have
\begin{align}
\Pr\left( \frac{1}{n} d_H(W_{\ell}^n, [X_{\ell}^n]_{\overline{{\cal X}}_\ell}) \ge \beta \right) 
 \le \frac{1}{\beta} \left[ \left( 1 + \frac{|{\cal X}_{{\cal A}^c}|}{q} \right) \varepsilon_n + \mu_n \right].
\end{align}
Then, by Lemma \ref{lemma:boosting}, there exists a code $(\kappa_n^{(\ell)},\tau_n^{(\ell)})$ of size $2^{n\delta}$ such that 
\begin{align}
\Pr\left( \tau_n^{(\ell)}( \kappa_n^{(\ell)}( [X^n_\ell]_{\overline{{\cal X}}_\ell}), W_\ell^n) \neq  [X^n_\ell]_{\overline{{\cal X}}_\ell} \right) \le
  \frac{1}{\beta} \left[ \left( 1 + \frac{|{\cal X}_{{\cal A}^c}|}{q} \right) \varepsilon_n + \mu_n \right] + \nu_n(\beta,|{\cal X}_\ell|) 2^{-n\delta}.
   \label{eq:error-probability-of-modified-code}
\end{align}
By taking $\delta > 0$ appropriately compared to $\beta > 0$, the error probability in the right hand side of \eqref{eq:error-probability-of-modified-code}
converges to $0$ for every $\ell \in {\cal A}$ and ${\cal A} \in \overline{{\cal L}}$. This means that, if each encoder sends additional message of rate $\delta$,
then the modified decoder can reproduce $[X^n_{\cal L}]_{\overline{{\cal X}}_{\cal L}}$ with vanishing error probability. Thus, 
$(R_\ell + \delta : \ell \in {\cal L}) \in {\cal R}(\bm{X}_{\cal L}|\bm{f}_{\overline{{\cal X}}_{\cal L}})$. Since $\beta > 0$ can be arbitrarily small, and 
we can make $\delta > 0$ arbitrarily small accordingly, we have 
\begin{align}
{\cal R}(\bm{X}_{\cal L}| \bm{f}) \subseteq {\cal R}(\bm{X}_{\cal L}|\bm{f}_{\overline{{\cal X}}_{\cal L}}).
\end{align}
We also have \eqref{theorem:main-semi} since the modified code can also compute $\bm{f}$ with vanishing error probability. 
\qed

\subsection{Proof of Achievability Part of Theorem \ref{theorem:general}}

The proof of the achievability part, i.e., $\supseteq$ part of \eqref{theorem:main-informative},
is essentially the same as \cite{KuzWat16}.
First, we claim that, given $[\bm{x}_\ell]_{\overline{{\cal X}}_\ell}$ and $P_{\bm{x}_\ell}$ of a sequence $\bm{x}_\ell \in {\cal X}_\ell^n$,
we can construct a sequence $\hat{\bm{x}}_\ell \in {\cal X}_\ell^n$ satisfying $[\hat{\bm{x}}_\ell]_{\overline{{\cal X}}_\ell} = [\bm{x}_\ell]_{\overline{{\cal X}}_\ell}$
and $\hat{\bm{x}}_\ell = \sigma_\ell(\bm{x}_\ell)$ for some permutation $\sigma_\ell$ on $[1:n]$. Indeed, we can construct 
$\hat{\bm{x}}_\ell = \hat{\bm{x}}_\ell([\bm{x}_\ell]_{\overline{{\cal X}}_\ell}, P_{\bm{x}_\ell})$ as follows. From $[\bm{x}_\ell]_{\overline{{\cal X}}_\ell}$, we can determine
a partition $\{ {\cal I}_{\cal C} : {\cal C} \in \overline{{\cal X}}_\ell \}$ of $[1:n]$ as 
\begin{align}
{\cal I}_{\cal C} := \{ i \in [1:n] : [x_{\ell,i}]_{\overline{{\cal X}}_\ell} = {\cal C} \},~~~{\cal C} \in \overline{{\cal X}}_\ell.
\end{align}
Then, for given $P_{\bm{x}_\ell}$, we can divide each ${\cal I}_{\cal C}$ into a partition $\{ {\cal I}_a : a \in {\cal C} \}$ so that\footnote{Although
there are several partitions which satisfy \eqref{eq:condition-on-partition}, the choice of a partition does not affect the argument; we may choose a 
partition so that, for $a,\hat{a} \in {\cal C}$ satisfying $a < \hat{a}$, $i \in {\cal I}_a$ and $j \in {\cal I}_{\hat{a}}$ imply $i < j$.}
\begin{align} \label{eq:condition-on-partition}
|{\cal I}_a| = n P_{\bm{x}_\ell}(a),~~~\forall a \in {\cal C} \subseteq {\cal X}_\ell.
\end{align}
Note that $\{ {\cal I}_a : a \in {\cal X}_\ell \}$ is also a partition of $[1:n]$; i.e., for each $i \in [1:n]$, there exists only one $\hat{x}_{\ell,i} \in {\cal X}_\ell$
such that $i \in {\cal I}_{\hat{x}_{\ell,i}}$. Then, it is not hard to see that $\hat{\bm{x}}_\ell = (\hat{x}_{\ell,1},\ldots,\hat{x}_{\ell,n})$ satisfies the desired property. 

Now, suppose that we are given a code $\hat{\Phi}_n = (\hat{\varphi}_n^{(1)},\ldots,\hat{\varphi}_n^{(L)},\hat{\psi}_n)$
for computing $[X^n_{{\cal L}}]_{\overline{{\cal X}}_{\cal L}} = f_{n,\overline{{\cal X}}_{\cal L}}(X^n_{\cal L})$
with error probability $\varepsilon_n$. From this code, we shall construct a code for computing $f_n$ as follows. 
In the new code, upon observing $X^n_\ell = \bm{x}_\ell$, the $\ell$th encoder sends the marginal type $P_{\bm{x}_\ell}$ of $\bm{x}_\ell$
by using $|{\cal X}_\ell | \log (n+1)$ bits in addition to the codeword $\hat{\varphi}_n^{(\ell)}(\bm{x}_\ell)$ of the original code. 
Suppose that the decoder obtained $[\bm{x}_\ell]_{\overline{{\cal X}}_\ell}$ from the codewords
$(\hat{\varphi}_n^{(\ell)}(\bm{x}_\ell) : \ell \in {\cal L})$. Then, since $P_{\bm{x}_\ell}$ is also sent from the $\ell$th encoder, 
for each $\ell \in {\cal L}$, the decoder can construct a sequence $\hat{\bm{x}}_\ell = \hat{\bm{x}}_\ell([\bm{x}_\ell]_{\overline{{\cal X}}_\ell}, P_{\bm{x}_\ell})$
satisfying $[\hat{\bm{x}}_\ell]_{\overline{{\cal X}}_\ell} = [\bm{x}_\ell]_{\overline{{\cal X}}_\ell}$ and
$\hat{\bm{x}}_\ell = \sigma_\ell(\bm{x}_\ell)$ for some permutation $\sigma_\ell$ on $[1:n]$ as shown above. 
Since $f_n$ is $\overline{{\cal X}}_{\cal A}$-informative for every ${\cal A} \in \overline{{\cal L}}$, we have
\begin{align}
f_n(\bm{x}_1,\bm{x}_2,\ldots,\bm{x}_L)
&= f_n(\hat{\bm{x}}_1,\bm{x}_2,\ldots,\bm{x}_L) \\
&= f_n(\hat{\bm{x}}_1,\hat{\bm{x}}_2,\ldots,\bm{x}_L) \\
&~~\vdots \\
&= f_n(\hat{\bm{x}}_1,\hat{\bm{x}}_2,\ldots,\hat{\bm{x}}_L).
\end{align}
This implies that the decoder can compute $f_n(\bm{x}_{\cal L})$ with error probability $\varepsilon_n$, and thus we have
\begin{align}
{\cal R}(\bm{X}_{\cal L} | \bm{f}_{\overline{{\cal X}}_{\cal L}}) \subseteq {\cal R}(\bm{X}_{\cal L} | \bm{f}).
\end{align}

\qed

\section{Two Terminals}
\label{section:two-terminal}

In this section, we apply Theorem \ref{theorem:general} for the two terminal setting. 

\subsection{Han-Kobayashi's Dichotomy}

In \cite{HanKob87}, in order to classify symbol-wise functions by whether the function computation regions
coincide with the Slepian-Wolf region, Han and Kobayashi introduced the following class of functions (see also \cite{KuzWat15}).
\begin{definition}[HK Function]
A function $f_n$ is called a Han-Kobayashi (HK) function if $f_n$ is a symbol-wise function defined by
some $f: {\cal X}_1 \times {\cal X}_2 \to {\cal V}$ such that 
\begin{enumerate}
\item \label{HK-condition-1}
for every $a \neq a^\prime$ in ${\cal X}_1$, the functions $f(a,\cdot)$ and $f(a^\prime,\cdot)$ are distinct,

\item \label{HK-condition-2}
for every $b \neq b^\prime$ in ${\cal X}_2$, the functions $f(\cdot,b)$ and $f(\cdot,b^\prime)$ are distinct, and

\item \label{HK-condition-3}
$f(a,b) \neq f(a^\prime,b^\prime)$ whenever $a \neq a^\prime$ and $b \neq b^\prime$.
\end{enumerate} 
\end{definition}

In Proposition \ref{proposition:informative-symbol-wise}, 
we have seen that the symbol-wise function is $\overline{{\cal X}}_{1,f}$-informative
and $\overline{{\cal X}}_{2,f}$-informative for partitions induced by $f$. It is not difficult to verify the following.
\begin{proposition} \label{proposition:HK-property-1}
If symbol-wise function $f^n$ satisfies
Conditions \ref{HK-condition-1} and \ref{HK-condition-2} of the HK function, respectively, then
it is ${\cal X}_1$-informative and ${\cal X}_2$-informative, respectively.\footnote{By a slight abuse of terminology, we say ${\cal X}_\ell$-informative
to mean $\overline{{\cal X}}_\ell$-informative for the finest partition $\overline{{\cal X}}_\ell = \{ \{ x_\ell \} : x_\ell \in {\cal X}_\ell \}$.}
\end{proposition}

We can find that Condition \ref{HK-condition-3} of the HK function can be rephrased as follows:
for every $v \in {\cal V}$, the inverse image satisfies either $f^{-1}(v) \subseteq {\cal X}_1 \times \{x_2\}$ for some $x_2 \in {\cal X}_2$
or $f^{-1}(v) \subseteq \{x_1\} \times {\cal X}_2$ for some $x_1 \in {\cal X}_1$. Thus, for a given correlated random variables $(X_1,X_2)$ 
on ${\cal X}_1 \times {\cal X}_2$, we have $X_1 \markov f(X_1,X_2) \markov X_2$, which implies the following.

\begin{proposition} \label{proposition:HK-property-2}
If symbol-wise function $\bm{f} = \{f_n\}_{n=1}^\infty$ satisfies Conditions \ref{HK-condition-3} of the HK function and $(\bm{X}_1,\bm{X}_2)$ is an i.i.d. source,
then $\bm{f}$ induces conditional independence for $((\bm{X}_1,\bm{X}_2), \{\{1\},\{2\} \})$.
\end{proposition}

By noting Propositions \ref{proposition:HK-property-1} and \ref{proposition:HK-property-2}, we can recover the sufficiency part of 
Han-Kobayashi's dichotomy theorem as a corollary of Theorem \ref{theorem:general}.
\begin{corollary}[\cite{HanKob87}]
If $\bm{f} = \{f_n\}_{n=1}^\infty$ is an HK function, then, for any i.i.d. sources satisfying the positivity condition,
we have ${\cal R}(\bm{X}_1,\bm{X}_2|\bm{f}) = {\cal R}_{\san{SW}}(\bm{X}_1,\bm{X}_2)$.
\end{corollary}

\subsection{Kuzuoka-Watanabe's Dichotomy}

In \cite{KuzWat15}, 
the class of joint sensitive functions was introduced.
\begin{definition}
A function $f_n: {\cal X}_1^n \times {\cal X}_2^n \to {\cal Z}_n$ is said to be {\em jointly sensitive}
if $f_n(\bm{x}_1,\bm{x}_2) \neq f_n(\hat{\bm{x}}_1,\hat{\bm{x}}_2)$ whenever $\bm{x}_1 \neq \hat{\bm{x}}_1$
and $\bm{x}_2 \neq \hat{\bm{x}}_2$.
\end{definition}

By a similar reasoning as Proposition \ref{proposition:HK-property-2},\footnote{Instead of one symbol function $f$, we consider 
the same argument for $f_n$.} we have the following property of jointly sensitive functions.
\begin{proposition} \label{proposition:property-joint-sensitivity}
If $\bm{f} = \{f_n \}_{n=1}^\infty$ is jointly sensitive, then, for any given source $(\bm{X}_1,\bm{X}_2)$ (not necessarily i.i.d.), 
$\bm{f}$ induces conditional independence for $((\bm{X}_1,\bm{X}_2), \{\{1\},\{2\} \})$.
\end{proposition}

By noting Proposition \ref{proposition:property-joint-sensitivity}, Theorem \ref{theorem:general} implies the following.
\begin{corollary} \label{corollary:jointy-sensitivity}
If $\bm{f}$ is $\overline{{\cal X}}_1$-informative and $\overline{{\cal X}}_2$-informative for a given partition $\overline{{\cal X}}_{\cal L} = (\overline{{\cal X}}_1,\overline{{\cal X}}_2)$
and $\bm{f}$ is jointly sensitive, then, for any smooth source $(\bm{X}_1,\bm{X}_2)$, we have
${\cal R}(\bm{X}_1,\bm{X}_2|\bm{f}) = {\cal R}(\bm{X}_1,\bm{X}_2|\bm{f}_{\overline{{\cal X}}_{\cal L}})$.
\end{corollary}

In general, a function being HK function does not imply the same function being jointly sensitive (cf.~\cite[Proposition 1]{KuzWat15}). 
Because of this fact, Han-Kobayashi's dichotomy theorem is not valid for the class of smooth sources, 
and an alternative dichotomy theorem was derived in \cite{KuzWat15}; the sufficiency part of the dichotomy theorem in \cite{KuzWat15}
can be recovered as a special case of Corollary \ref{corollary:jointy-sensitivity}.
\begin{corollary}[\cite{KuzWat15}]
If $\bm{f}$ is HK function and jointly sensitive, then, for any smooth sources, 
we have ${\cal R}(\bm{X}_1,\bm{X}_2|\bm{f}) = {\cal R}_{\san{SW}}(\bm{X}_1,\bm{X}_2)$.\footnote{In \cite{KuzWat15},
the dichotomy theorem was stated in terms of ``totally sensitive function"; however, a function being HK and jointly sensitive is
equivalent to HK and totally sensitive.}
\end{corollary} 

\subsection{Mixed Sources} \label{subsec:mixture}

When a source has memory, the statement of Proposition \ref{proposition:HK-property-2} may not hold in general. 
Thus, it was not clear if a function being HK (but not jointly sensitive) implies ${\cal R}(\bm{X}_1,\bm{X}_2|\bm{f}) = {\cal R}_{\san{SW}}(\bm{X}_1,\bm{X}_2)$
when the source has memory. Here, we show that this claim holds for a mixture of i.i.d. sources.

Let $(\bm{X}_1,\bm{X}_2) = \{ (X_1^n,X_2^n) \}_{n=1}^\infty$ be a mixture of two i.i.d. sources with positivity condition such that 
\begin{align}
P_{X_1^n X_2^n}(\bm{x}_1,\bm{x}_2) = \Pr( A=0) P_{X_1 X_2,0}^n(\bm{x}_1,\bm{x}_2) + \Pr(A=1) P_{X_1 X_2,1}^n(\bm{x}_1,\bm{x}_2)
\end{align}
for some random variable $A$ taking values in $\{0,1\}$. Let $f_n:{\cal X}_1^n \times {\cal X}_2 \to {\cal V}^n$ be an HK function. Suppose that $P_{V,0} \neq P_{V,1}$, where 
$P_{V,a}$ is the distribution of $V = f(X_1,X_2)$ under $P_{X_1X_2,a}$. Then, there exists an estimator $\gamma_n$ such that the estimation error probability
$\Pr( A \neq \gamma_n(V^n))$ vanishes asymptotically. Thus, by taking $S_n = (A,V^n)$, the function $\bm{f} = \{f_n \}_{n=1}^\infty$ induces 
conditional independence for $((\bm{X}_1,\bm{X}_2), \{ \{1\}, \{2\} \})$. Consequently, Theorem \ref{theorem:general} and Proposition \ref{proposition:HK-property-1} imply 
${\cal R}(\bm{X}_1,\bm{X}_2|\bm{f}) = {\cal R}_{\san{SW}}(\bm{X}_1,\bm{X}_2)$.

\section{More Than Two Terminals with I.I.D. Sources}
\label{section:multi-terminal-iid}


In \cite{HanKob87}, Han-Kobayashi considered classification of functions for more than two terminals; 
let ${\cal F}_{\san{SW}}^{\san{iid}}$ be the set of all functions $f$ such that 
the symbol-wise function induced by $f$ satisfies ${\cal R}(\bm{X}_{\cal L}|\bm{f}) = {\cal R}_{\san{SW}}(\bm{X}_{\cal L})$ 
for any i.i.d. sources with positivity condition.
They derived a necessary condition and sufficient conditions such that a given function belongs to ${\cal F}_{\san{SW}}^{\san{iid}}$.
In this section, as an application of the general results in Section \ref{section:general-results}, we derive a novel 
sufficient condition that strictly subsumes the sufficient conditions in \cite{HanKob87}. 

First, we shall review the necessary condition and the sufficient conditions derived in \cite{HanKob87}.
For that purpose, it is convenient to introduce some ``geometrical" notation of functions.

\begin{definition}[Projection] \label{definition:projection}
For $f: {\cal X}_{\cal L} \to {\cal V}$ and a subset ${\cal A} \subseteq {\cal L}$, we define projected function
$f_{\cal A} : {\cal X}_{\cal A} \to {\cal V}^{|{\cal X}_{{\cal A}^c}|}$ by\footnote{For ${\cal A} = {\cal L}$, note that $f_{\cal L} = f$.}
\begin{align}
f_{\cal A}(x_{\cal A}) = \big( f(x_{\cal A}, x_{{\cal A}^c}) : x_{{\cal A}^c} \in {\cal X}_{{\cal A}^c} \big). 
\end{align}
\end{definition}

\begin{definition}[Span] \label{definition:span}
For $f: {\cal X}_{\cal L} \to {\cal V}$ and $v \in {\cal V}$, we define $\mathtt{span} f^{-1}(v)$ as the minimal subset ${\cal B} \subseteq {\cal L}$
such that there exists $x_{{\cal B}^c} \in {\cal X}_{{\cal B}^c}$ satisfying 
$f^{-1}(v) \subseteq {\cal X}_{{\cal B}} \times \{ x_{{\cal B}^c} \}$.
In particular, when $|f^{-1}(v)|\le1$, then $\mathtt{span}f^{-1}(v) = \emptyset$.
For a subset ${\cal A} \subseteq {\cal L}$, $\mathtt{span}f_{\cal A}^{-1}(\bm{v})$ is defined similarly by replacing $f$ with 
projected function $f_{\cal A} :{\cal X}_{\cal A} \to {\cal V}^{|{\cal X}_{{\cal A}^c}|}$.
\end{definition}

Essentially, $\mathtt{span} f^{-1}(v)$ is the set of coordinates $\ell$ such that the value of $x_\ell$ cannot be uniquely determined from $v$.
Let us verify Definition \ref{definition:projection} and Definition \ref{definition:span} with some examples.

\begin{figure}[tb]
\centering{
\includegraphics[width=0.13\textwidth, bb=0 0 82 79]{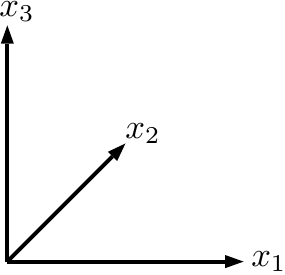}
\caption{Coordinates.}
\label{Fig:coordinates}
}
\end{figure}

\begin{figure}[!t]
  \centering
  \subfloat[][]{\includegraphics[width=.15\textwidth, bb=0 0 129 138]{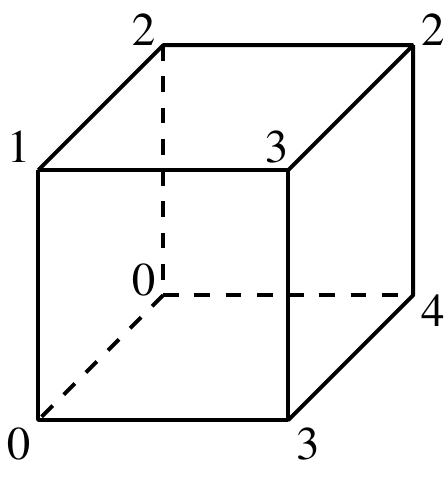} \label{Fig:pattern-A}}  \quad
  \subfloat[][]{\includegraphics[width=.15\textwidth, bb=0 0 129 138]{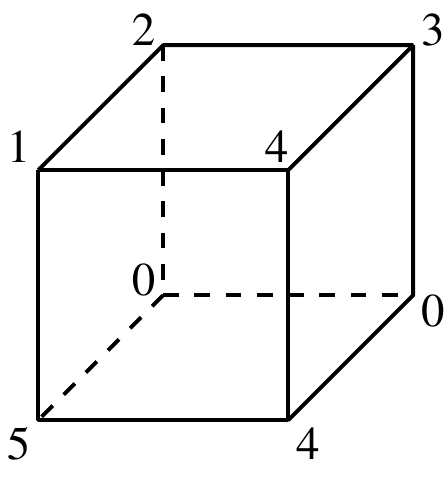} \label{Fig:pattern-B}}  \quad
  \subfloat[][]{\includegraphics[width=.15\textwidth, bb=0 0 129 138]{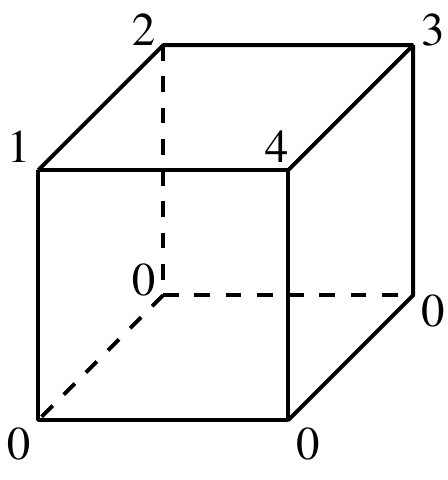} \label{Fig:pattern-C}}  \\
  \subfloat[][]{\includegraphics[width=.15\textwidth, bb=0 0 129 138]{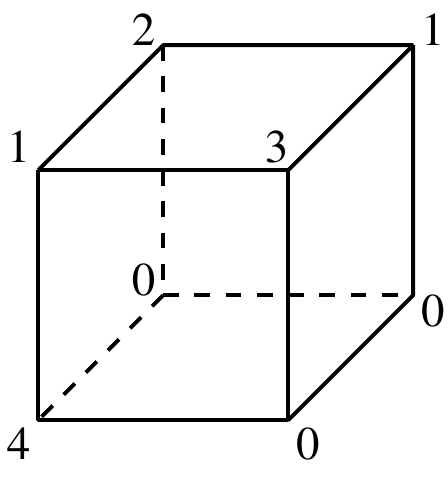} \label{Fig:pattern-D}} \quad \hspace{.15\textwidth} \quad
  \subfloat[][]{\includegraphics[width=.15\textwidth, bb=0 0 129 138]{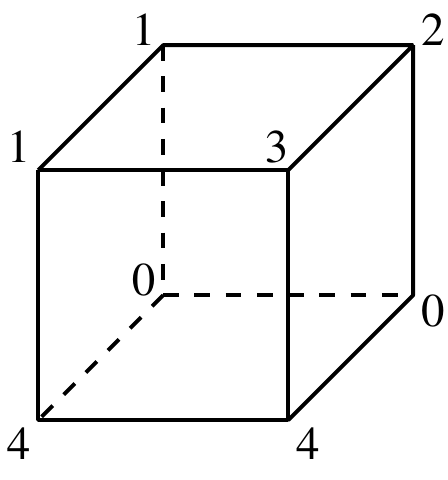} \label{Fig:pattern-E}} \\
  \subfloat[][]{\includegraphics[width=.15\textwidth, bb=0 0 129 138]{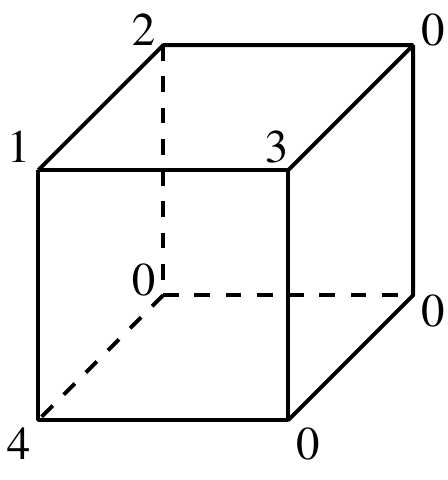} \label{Fig:pattern-F}} \quad \hspace{.15\textwidth} \quad
  \subfloat[][]{\includegraphics[width=.15\textwidth, bb=0 0 129 138]{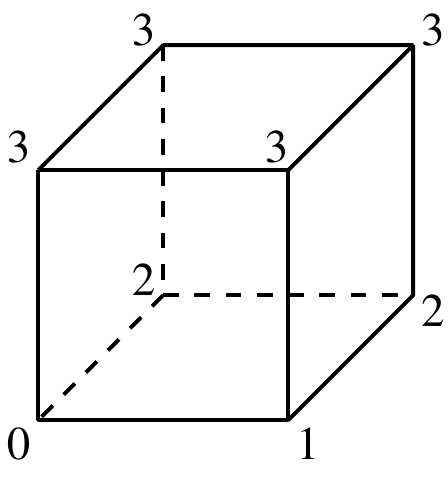} \label{Fig:pattern-Recursion}} \\
  \subfloat[][]{\includegraphics[width=.15\textwidth, bb=0 0 129 138]{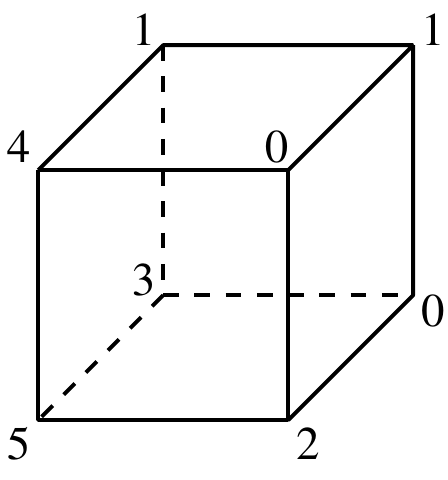} \label{Fig:pattern-H}} \quad
  \subfloat[][]{\includegraphics[width=.15\textwidth, bb=0 0 129 138]{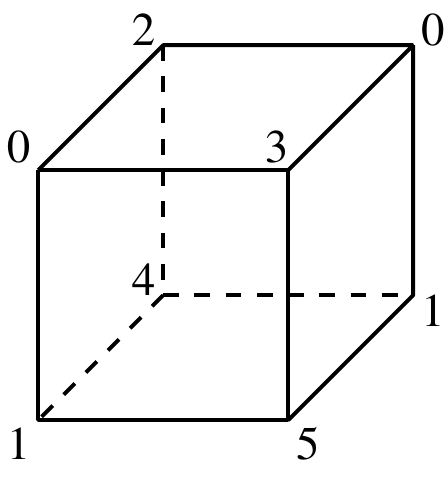} \label{Fig:pattern-I}} \quad 
  \subfloat[][]{\includegraphics[width=.15\textwidth, bb=0 0 129 138]{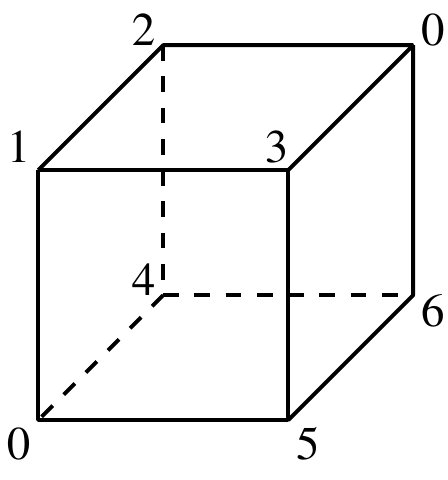} \label{Fig:pattern-J}}
  \caption{Representative patterns for $f(x_1,x_2,x_3)$.}
  \label{Fig:pattern-three-terminal}
\end{figure}

\begin{example} \label{example:geometrical-notation}
Let us consider three terminal function $f(x_1,x_2,x_3)$, where values of $(x_1,x_2,x_3)$
are ordered as in Fig.~\ref{Fig:coordinates}.\footnote{This method of describing functions was introduced in \cite{HanKob87}.}
For the function in Fig.~\ref{Fig:pattern-three-terminal}(a), $f_{\{3\}}(0) = (0,0,3,4)$ and $f_{\{3\}}(1) = (1,2,3,2)$.
For the same function, $\mathtt{span} f^{-1}(1) = \emptyset$ and $\mathtt{span} f^{-1}(0) = \{2\}$, respectively. 
For the function in Fig.~\ref{Fig:pattern-three-terminal}(f), $\mathtt{span} f^{-1}(0) = \{ 1,2,3\}$.
\end{example}

Now, we are ready to review the conditions derived by Han and Kobayashi \cite{HanKob87}, where we
rephrase their conditions by using the notations introduced above.

\begin{proposition}[\cite{HanKob87}; Necessary Condition] \label{propositioon:necessary-HK}
A function $f:{\cal X}_{\cal L} \to {\cal V}$ belongs to ${\cal F}_{\san{SW}}^{\san{iid}}$ only if
the following conditions holds: for every ${\cal A} \subseteq {\cal L}$, 
\begin{align}
f_{\cal A}(x_{\cal A}) \neq f_{\cal A}(\hat{x}_{\cal A})
\end{align}
holds for any pair $x_{\cal A}, \hat{x}_{\cal A} \in {\cal X}_{\cal A}$ satisfying
$x_\ell \neq \hat{x}_\ell$ for every $\ell \in {\cal A}$. 
\end{proposition}

\begin{proposition}[\cite{HanKob87}; Sufficient Condition] \label{proposition:HK-sufficiency-1}
A function $f:{\cal X}_{\cal L} \to {\cal V}$ belongs to ${\cal F}_{\san{SW}}^{\san{iid}}$ if
the following two conditions hold:
\begin{enumerate}
\item \label{condition:HK-sufficiency-1-1}
For any $v \in {\cal V}$, the inverse image satisfies $|\mathtt{span} f^{-1}(v)|\le1$.

\item \label{condition:HK-sufficiency-1-2}
For arbitrary $\ell \in {\cal L}$, the projected function $f_{{\cal L}\backslash \{\ell \}}$ is injective.
\end{enumerate}
\end{proposition}

\begin{proposition}[\cite{HanKob87}; Sufficient Condition] \label{proposition:HK-sufficiency-2}
A function $f:{\cal X}_{\cal L}\to {\cal V}$ belongs to ${\cal F}_{\san{SW}}^{\san{iid}}$ if,
in addition to Condition \ref{condition:HK-sufficiency-1-2} in Proposition \ref{proposition:HK-sufficiency-1}, the following condition holds:
\begin{enumerate}
\renewcommand\theenumi{\arabic{enumi}'}
\item \label{condition:HK-sufficiency-2-1}
\begin{align}
\bigcup_{v \in {\cal V}} \mathtt{span} f^{-1}(v) \subsetneq {\cal L}.
\end{align}
\end{enumerate}
\end{proposition}

\textchange{In the following example, we examine if functions in Fig.~\ref{Fig:pattern-three-terminal} satisfy the conditions in Propositions \ref{propositioon:necessary-HK}, 
\ref{proposition:HK-sufficiency-1}, and \ref{proposition:HK-sufficiency-2}, which is summarized in Table \ref{table:sufficient-necessary}.\footnote{All the functions in Fig.~\ref{Fig:pattern-three-terminal} except 
(g) and (h) are the same as those in \cite[Fig.~4]{HanKob87} and \cite[Fig.~5]{HanKob87}.} 
In the rightmost column of the table, we also described if we can decide $f \in {\cal F}_{\san{SW}}^{\san{iid}}$ or not by using
Propositions \ref{propositioon:necessary-HK}, \ref{proposition:HK-sufficiency-1}, and \ref{proposition:HK-sufficiency-2}, and Theorem \ref{theorem:condition-symbol-wise-function-iid} shown later in this section. 
}

\begin{example}
All the functions in Fig.~\ref{Fig:pattern-three-terminal}(a)-(h)
satisfy the necessary condition in Proposition \ref{propositioon:necessary-HK}; 
the functions in Fig.~\ref{Fig:pattern-three-terminal}(i) and (j)
violate the necessary condition in Proposition \ref{propositioon:necessary-HK}. 
From Proposition \ref{propositioon:necessary-HK}, we can determine that the functions in (i) and (j) do not belong to ${\cal F}_{\san{SW}}^{\san{iid}}$;
on the other hand, from Proposition \ref{proposition:HK-sufficiency-1} and Proposition \ref{proposition:HK-sufficiency-2}, we can determine that
the functions in (a), (b), and (d) belong to ${\cal F}_{\san{SW}}^{\san{iid}}$. In \cite{HanKob87}, it is claimed that, by some inspection, 
the function in (c) is found to belong to ${\cal F}_{\san{SW}}^{\san{iid}}$; later in Example \ref{example:pattern-C}, we will verify this fact from more general sufficient condition.
\begin{table}[tb]
\centering{
\caption{Necessary/Sufficient Conditions for Functions in Fig.~\ref{Fig:pattern-three-terminal}}
\begin{tabular}{|c|c|c|c|c|c|} \hline \label{table:sufficient-necessary} 
 $f(x_1,x_2,x_3)$ & Neces. Cond. (Prop. \ref{propositioon:necessary-HK}) & Suff. Cond. \ref{condition:HK-sufficiency-1-1} (Prop. \ref{proposition:HK-sufficiency-1}) & 
  Suff. Cond. \ref{condition:HK-sufficiency-2-1} (Prop. \ref{proposition:HK-sufficiency-2}) & Suff. Cond. \ref{condition:HK-sufficiency-1-2} (Prop. \ref{proposition:HK-sufficiency-1}) & $f \in {\cal F}_{\san{SW}}^{\san{iid}}$ \\\hline
 (a) & \checkmark & \checkmark &  & \checkmark & \checkmark (Prop. \ref{proposition:HK-sufficiency-1}) \\ \hline
 (b) & \checkmark & \checkmark &  &\checkmark & \checkmark (Prop. \ref{proposition:HK-sufficiency-1}) \\ \hline
 (c) & \checkmark &  & \checkmark & &  \checkmark (Thm. \ref{theorem:condition-symbol-wise-function-iid} ) \\ \hline
 (d) & \checkmark &  & \checkmark & \checkmark & \checkmark (Prop. \ref{proposition:HK-sufficiency-2}) \\ \hline
 (e) & \checkmark & \checkmark &  & & \checkmark (Thm. \ref{theorem:condition-symbol-wise-function-iid}) \\ \hline
 (f)  & \checkmark &  &  & \checkmark & \\ \hline
 (g)  & \checkmark  &  & \checkmark & & \checkmark (Thm. \ref{theorem:condition-symbol-wise-function-iid}) \\ \hline
 (h) & \checkmark & & & \checkmark & \checkmark (Thm. \ref{theorem:condition-symbol-wise-function-iid}) \\ \hline
 (i)  & & \checkmark &  &  & $\times$ (Prop. \ref{propositioon:necessary-HK}) \\ \hline
 (j)  &  &  &  & \checkmark & $\times$ (Prop. \ref{propositioon:necessary-HK}) \\ \hline
\end{tabular}
}
\end{table}
\end{example}

The conditions in Propositions \ref{proposition:HK-sufficiency-1} and \ref{proposition:HK-sufficiency-2} can be rephrased 
in terms of the general framework in Section \ref{section:general-results} as follows. When function $f$ satisfies 
Conditions \ref{condition:HK-sufficiency-1-1} and \ref{condition:HK-sufficiency-1-2}
of Proposition \ref{proposition:HK-sufficiency-1}, then it induces conditional independence for $(\bm{X}_{\cal L},\overline{{\cal L}})$ with the finest 
partition $\overline{{\cal L}} \equiv {\cal L}$ and it is semi ${\cal X}_\ell$-informative for every $\ell \in {\cal L}$. Thus, Corollary \ref{corollary:recursion-SW} with $k=1$
implies $f \in {\cal F}_{\san{SW}}^{\san{iid}}$. 
On the other hand, when function $f$ satisfies Condition \ref{condition:HK-sufficiency-1-2} of Proposition \ref{proposition:HK-sufficiency-1}
and Condition \ref{condition:HK-sufficiency-2-1} of Proposition \ref{proposition:HK-sufficiency-2}, then it induces conditional independence for
$(\bm{X}_{\cal L},\overline{{\cal L}})$ with partiton $\overline{{\cal L}} = \{ \{\ell_0\}, {\cal L}\backslash\{ \ell_0\} \}$ for some fixed $\ell_0 \in {\cal L}$ and it is
semi ${\cal X}_{{\cal L}\backslash\{\ell_0\}}$-informative and semi ${\cal X}_{\ell_0}$-informative. Thus, Corollary \ref{corollary:recursion-SW} with $k=1$
again implies $f \in {\cal F}_{\san{SW}}^{\san{iid}}$.
In fact, Condition \ref{condition:HK-sufficiency-1-2} of Proposition \ref{proposition:HK-sufficiency-1} is quite strong since it implies 
that $f$ is semi ${\cal X}_{{\cal L}\backslash\{\ell\}}$-informative for every $\ell \in {\cal L}$.

As we discussed above, Propositions \ref{proposition:HK-sufficiency-1} and \ref{proposition:HK-sufficiency-2}
are implied as special cases of Corollary \ref{corollary:recursion-SW}. In order to leverage Corollary \ref{corollary:recursion-SW}
in a full generality, let us rephrase conditional independence structure and informative structure in terms of the basis function
$f:{\cal X}_{\cal L} \to {\cal V}$, which will be proved at the end of this section.

\begin{theorem} \label{theorem:condition-symbol-wise-function-iid}
For a given nontrivial partition $\overline{{\cal L}}$ of ${\cal L}$ and tuple of partitions $\overline{{\cal X}}_{\cal L} = (\overline{{\cal X}}_\ell : \ell \in {\cal L})$,
suppose that $f:{\cal X}_{\cal L} \to {\cal V}$ 
satisfies the following condition:
\begin{enumerate}
\item \label{condition:new-1}
For every $v \in {\cal V}$, the inverse image satisfies $\mathtt{span} f^{-1}(v) \subseteq {\cal A}$ for some ${\cal A} \in \overline{{\cal L}}$.

\end{enumerate}
Then, for any i.i.d. source $\bm{X}_{\cal L}$, 
the symbol-wise function $f^n$ induces conditional independence for $(\bm{X}_{\cal L}, \overline{{\cal L}})$.
On the other hand, suppose that
$f$ satisfies the following condition:
\begin{enumerate}
\setcounter{enumi}{1}
\item \label{condition:new-2}
For every ${\cal A} \in \overline{{\cal L}}$ and every $x_{\cal A}, \hat{x}_{\cal A} \in {\cal X}_{\cal A}$ with 
$[x_{\cal A}]_{\overline{{\cal X}}_{\cal A}} \neq [\hat{x}_{\cal A}]_{\overline{{\cal X}}_{\cal A}}$, the projected function satisfies
\begin{align}
f_{\cal A}(x_{\cal A}) \neq f_{\cal A}(\hat{x}_{\cal A}).
\end{align}
\end{enumerate}
Then, the symbol-wise function $f^n$ is semi $\overline{{\cal X}}_{\cal A}$-informative for every ${\cal A} \in \overline{{\cal L}}$. 
\end{theorem}

We can use Theorem \ref{theorem:condition-symbol-wise-function-iid} and Corollary \ref{corollary:recursion-SW} to verify if a given function $f$
is included in ${\cal F}_{\san{SW}}^{\san{iid}}$. More specifically, if a given function satisfies the two conditions in Theorem \ref{theorem:condition-symbol-wise-function-iid},
then we can use Corollary \ref{corollary:recursion-SW} to show that that function is included in ${\cal F}_{\san{SW}}^{\san{iid}}$.
We will illustrate the utility of this approach with some examples in the following.

\begin{example} \label{example:pattern-C}
Let us consider the function in Fig.~\ref{Fig:pattern-three-terminal}(c). For the partition $\overline{{\cal L}} = \{ \{1,2\}, \{3\} \}$,
since $\mathtt{span} f^{-1}(0) = \{1,2\}$ and $\mathtt{span} f^{-1}(v) = \emptyset$ for $v =1,\ldots,4$, 
we can verify that Condition \ref{condition:new-1} of Theorem \ref{theorem:condition-symbol-wise-function-iid} is satisfied.
Furthermore, for the finest partitions $\overline{{\cal X}}_i \equiv {\cal X}_i$ for $i=1,2,3$, since the projected functions $f_{\{1,2\}}$
and $f_{\{3\}}$ are both injective,
we can verify that Condition \ref{condition:new-2} is satisfied. Thus, 
the function in Fig.~\ref{Fig:pattern-three-terminal}(c) belongs to ${\cal F}_{\san{SW}}^{\san{iid}}$.
\end{example}

\begin{example} \label{example:pattern-E}
Let us consider the function in Fig.~\ref{Fig:pattern-three-terminal}(e). There are two manners to show that this function belongs to
${\cal F}_{\san{SW}}^{\san{iid}}$. The first manner is the same as Example \ref{example:pattern-C}: For the partition $\overline{{\cal L}} = \{ \{1,2\}, \{3\} \}$,
we can verify that Condition \ref{condition:new-1} of Theorem \ref{theorem:condition-symbol-wise-function-iid} is satisfied.
Furthermore, for the finest partitions $\overline{{\cal X}}_i \equiv {\cal X}_i$ for $i=1,2,3$, we can verify that 
Condition \ref{condition:new-2} is satisfied. Thus, it belongs to ${\cal F}_{\san{SW}}^{\san{iid}}$.
The second manner is as follows: For the partition $\overline{{\cal L}} = \{ \{1\}, \{2\}, \{3\} \}$, 
we can verify that Condition \ref{condition:new-1} of Theorem \ref{theorem:condition-symbol-wise-function-iid} is satisfied.
Furthermore, for the finest partitions $\overline{{\cal X}}_i \equiv {\cal X}_i$ for $i=1,2,3$, we can verify that 
Condition \ref{condition:new-2} is satisfied. Thus, it belongs to ${\cal F}_{\san{SW}}^{\san{iid}}$.
\end{example}

\begin{example}
Let us consider the function in Fig.~\ref{Fig:pattern-three-terminal}(h).
For the partition $\overline{{\cal L}} = \{ \{1\}, \{2,3\} \}$, since $\mathtt{span} f^{-1}(0) = \{2,3\}$, $\mathtt{span} f^{-1}(1) = \{1\}$,
and $\mathtt{span} f^{-1}(v)=\emptyset$ for $v=2,\ldots,5$,
we can verify that Condition \ref{condition:new-1} of Theorem \ref{theorem:condition-symbol-wise-function-iid} is satisfied.
Furthermore, for the finest partitions $\overline{{\cal X}}_i \equiv {\cal X}_i$ for $i=1,2,3$, we can verify that 
Condition \ref{condition:new-2} is satisfied. Thus, it belongs to ${\cal F}_{\san{SW}}^{\san{iid}}$.
\end{example}

For some examples, we need to use recursion of Corollary \ref{corollary:recursion-SW}.

\begin{example} \label{example:pattern-Recursion}
Let us consider the function in Fig.~\ref{Fig:pattern-three-terminal}(g). To prove that this function belongs to ${\cal F}_{\san{SW}}^{\san{iid}}$,
we need to use recursion of Corollary \ref{corollary:recursion-SW}. For the partition $\overline{{\cal L}}^{(1)} = \{ \{1,2\}, \{3\} \}$, 
we can verify that $f$ satisfies Condition \ref{condition:new-1} of Theorem \ref{theorem:condition-symbol-wise-function-iid}.
Furthermore, for the partitions $\overline{{\cal X}}^{(1)}_1 = \{ \{0,1\} \}$, $\overline{{\cal X}}^{(1)}_2 = \{ \{0\},\{1\} \}$, and $\overline{{\cal X}}^{(1)}_3 = \{\{0\}, \{1\}\}$,
we can verify that $f$ satisfies Condition \ref{condition:new-2} of Theorem \ref{theorem:condition-symbol-wise-function-iid}.
Thus, we have ${\cal R}(\bm{X}_{{\cal L}}|\bm{f}) = {\cal R}(\bm{X}_{{\cal L}} | \bm{f}, \bm{f}_{\overline{{\cal X}}_{\cal L}^{(1)}})$, where
$(\bm{f}, \bm{f}_{\overline{{\cal X}}_{\cal L}^{(1)}})$ is the sequence of symbol-wise functions that consists of the product of $f$ and local function $f_{\overline{{\cal X}}_{{\cal L}}^{(1)}}$.
In the recursion step, we shall show that  the product function $(f, f_{\overline{{\cal X}}_{{\cal L}}^{(1)}})$ belongs to ${\cal F}_{\san{SW}}^{\san{iid}}$.
Since $\overline{{\cal X}}_1$ is the trivial partition and $\overline{{\cal X}}_2$ and $\overline{{\cal X}}_3$ are the finest partition, 
$(f, f_{\overline{{\cal X}}_{{\cal L}}^{(1)}})$ is equivalent to $(f(x_1,x_2,x_3), x_2,x_3)$; see Fig.~\ref{Fig:example-Recursion-step}. 
Then, for the partition $\overline{{\cal L}}^{(2)} = \{ \{1\}, \{2\},\{3\} \}$,
we can verify that $(f, f_{\overline{{\cal X}}_{{\cal L}}^{(1)}})$ satisfies Condition \ref{condition:new-1} of Theorem \ref{theorem:condition-symbol-wise-function-iid}.
Furthermore, for the finest partitions $\overline{{\cal X}}^{(2)}_i \equiv {\cal X}_i$ for $i=1,2,3$, we can verify that 
$(f, f_{\overline{{\cal X}}_{{\cal L}}^{(1)}})$ satisfies Condition \ref{condition:new-2} of Theorem \ref{theorem:condition-symbol-wise-function-iid}. 
Thus, the function $(f,f_{\overline{{\cal X}}^{(1)}_{\cal L}})$ belongs to
${\cal F}_{\san{SW}}^{\san{iid}}$, which implies that $f$ belongs to ${\cal F}_{\san{SW}}^{\san{iid}}$.
\begin{figure}[tb]
\centering{
\includegraphics[width=0.23\textwidth, bb=0 0 188 141]{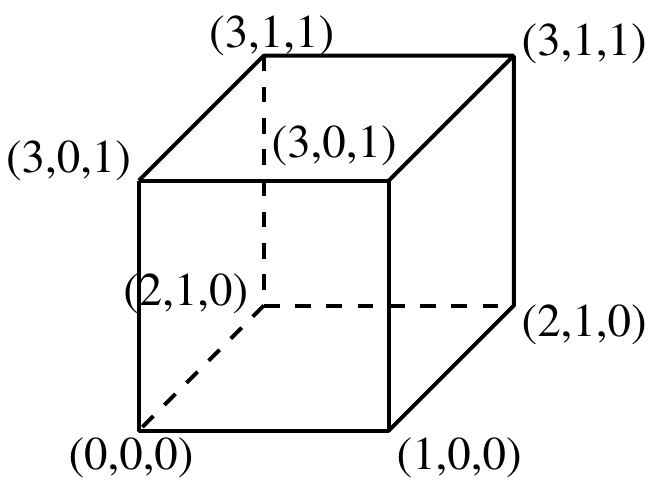}
\caption{The function in recursion step of Example \ref{example:pattern-Recursion}.}
\label{Fig:example-Recursion-step}
}
\end{figure}
\end{example}

\begin{example} \label{example:pattern-Recursion-3}
Let us consider the function in Fig.~\ref{Fig:pattern-recursion-3}(a), where ${\cal X}_1 = {\cal X}_2 = \{0,1\}$
and ${\cal X}_3 = \{0,1,2\}$. To prove that this function belongs to  ${\cal F}_{\san{SW}}^{\san{iid}}$,
we need to use recursion of Corollary \ref{corollary:recursion-SW}. For the partition $\overline{{\cal L}}^{(1)} = \{ \{1\}, \{2,3\} \}$, 
we can verify that $f$ satisfies Condition \ref{condition:new-1} of Theorem \ref{theorem:condition-symbol-wise-function-iid}.
Furthermore, for the partitions $\overline{{\cal X}}_1^{(1)} = \{ \{0\}, \{1\}\}$, $\overline{{\cal X}}_2^{(1)} = \{ \{0,1\} \}$, and 
$\overline{{\cal X}}_3^{(1)} = \{ \{0\}, \{1,2\}\}$, we can verify that $f$ satisfies Condition \ref{condition:new-2} of Theorem \ref{theorem:condition-symbol-wise-function-iid}.
Thus, we have ${\cal R}(\bm{X}_{{\cal L}}|\bm{f}) = {\cal R}(\bm{X}_{{\cal L}} | \bm{f}, \bm{f}_{\overline{{\cal X}}_{\cal L}^{(1)}})$.
In the recursion step, since $\overline{{\cal X}}_2^{(1)}$ is the trivial partition, 
$(f, f_{\overline{{\cal X}}_{{\cal L}}^{(1)}})$ is equivalent to $(f(x_1,x_2,x_3), x_1,[x_3]_{\overline{{\cal X}}_3^{(1)}})$; see Fig.~\ref{Fig:pattern-recursion-3}(b).
Thus, for the partition $\overline{{\cal L}}^{(2)} = \{ \{1\}, \{2\},\{3\} \}$,
we can verify that $(f, f_{\overline{{\cal X}}_{{\cal L}}^{(1)}})$ satisfies Condition \ref{condition:new-1} of Theorem \ref{theorem:condition-symbol-wise-function-iid}.
Furthermore, for the partitions $\overline{{\cal X}}^{(2)}_i = \{ \{0\}, \{1\} \}$ for $i=1,2$ and $\overline{{\cal X}}_3^{(2)} = \{ \{0\}, \{1\},\{2\}\}$, we can verify that 
$(f, f_{\overline{{\cal X}}_{{\cal L}}^{(1)}})$ satisfies Condition \ref{condition:new-2} of Theorem \ref{theorem:condition-symbol-wise-function-iid}. 
Thus, the function $(f,f_{\overline{{\cal X}}^{(1)}_{\cal L}})$ belongs to
${\cal F}_{\san{SW}}^{\san{iid}}$, which implies that $f$ belongs to ${\cal F}_{\san{SW}}^{\san{iid}}$.
\begin{figure}[!t]
  \centering
  \subfloat[][]{\includegraphics[width=0.17\textwidth, bb=0 0 129 210]{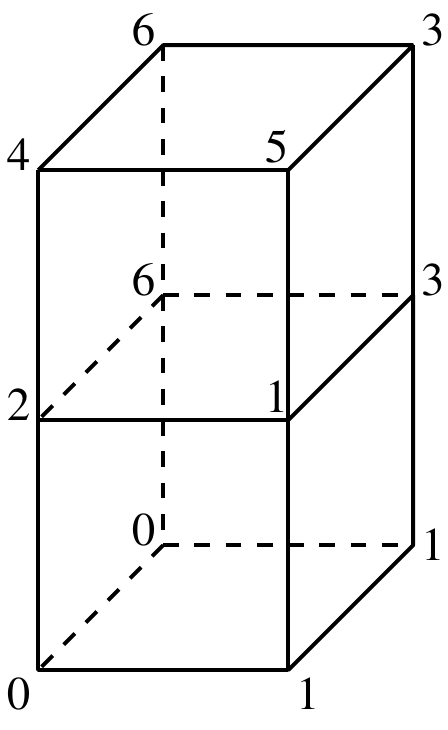} \label{Fig:pattern-Recursion3}}  \quad
  \subfloat[][]{\includegraphics[width=0.25\textwidth, bb=0 0 188 210]{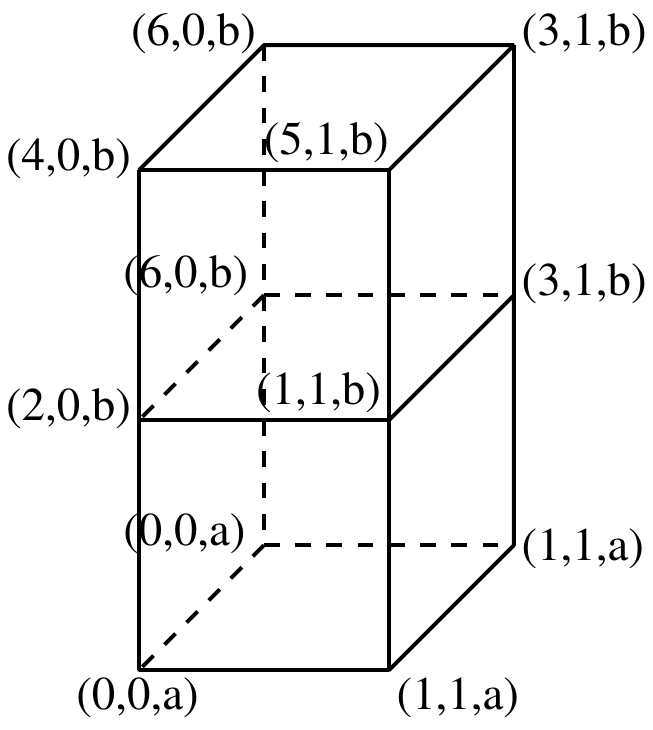} \label{Fig:pattern-Recursion4}}  
  \caption{(a) Description of $f(x_1,x_2,x_3)$ in Example \ref{example:pattern-Recursion-3}; (b) Description of the function
  in recursion step in the same example, where $a = \{0\}$ and $b = \{1,2\}$.}
  \label{Fig:pattern-recursion-3}
\end{figure}
\end{example}

So far, we have seen examples such that the classification problem can be solved by our approach.
However, there are still some cases such that the classification problem cannot be solved by our approach.
The following is one of such examples. 

\begin{example} \label{example:unclear-case}
Let us consider the function in Fig.~\ref{Fig:pattern-three-terminal}(f). In this case,
Condition \ref{condition:new-1} of Theorem \ref{theorem:condition-symbol-wise-function-iid} is not satisfied for any
nontrivial partition $\overline{{\cal L}}$ of $\{1,2,3\}$; in fact, $\mathtt{span} f^{-1}(0) = \{1,2,3\}$. 
Thus, we cannot determine whether $f$ belongs to ${\cal F}_{\san{SW}}^{\san{iid}}$
or not.
\end{example}

In the case of three terminals, a nontrivial partition $\overline{{\cal L}}$ inevitably include a singleton part.
In the case of more than three terminals, this is not the case. Even when the function induces conditional independence 
among non-singleton groups of parties, the classification problem can be solved,
which is illustrated in the following example.

\begin{example} \label{example:four-terminal}
Let us consider the function in Table \ref{table:Example-4A}.
For the partition $\overline{{\cal L}} = \{ \{1,2\}, \{3,4\}\}$, 
we can verify that Condition \ref{condition:new-1} of Theorem \ref{theorem:condition-symbol-wise-function-iid} is satisfied.
Furthermore, for the finest partitions $\overline{{\cal X}}_i = {\cal X}_i$ for $i=1,2,3,4$, we can verify that 
Condition \ref{condition:new-2} is satisfied. Thus, the function in Table \ref{table:Example-4A} belongs to ${\cal F}_{\san{SW}}^{\san{iid}}$.
\begin{table}[tb]
\centering{
\caption{Description of $f(x_1,x_2,x_3,x_4)$ in Example \ref{example:four-terminal}.}
\begin{tabular}{c|c|c|c|c|} \label{table:Example-4A}
 $(x_1,x_2)\setminus (x_3,x_4)$& 00 & 01& 10 & 11 \\\hline
 00 & 0 & 0 & 4 & 0 \\ \hline
 01 & 1 & 6 & 4 & 2 \\ \hline
 10 & 1 & 6 & 4 & 5 \\ \hline
 11 & 1 & 3 & 3 & 3 \\ \hline
\end{tabular}
}
\end{table}
\end{example}

\paragraph*{Proof of Theorem \ref{theorem:condition-symbol-wise-function-iid}} 

Since the source is i.i.d. and the function is symbol-wise, 
in order to prove that the function induces conditional independence, 
it suffices to show that,
for $X_{\cal L} \sim P_{X_{\cal L}}$ and $V = f(X_{\cal L})$, 
the joint distribution $P_{X_{\cal L} V}$ factorizes as
\begin{align} \label{eq:proof-symbol-wise-iid-factorize}
P_V \prod_{{\cal A} \in \overline{{\cal L}}} P_{X_{\cal A}|V}.
\end{align}
Since function $f$ satisfies Condition \ref{condition:new-1},
for each $v \in {\cal V}$, we have $f^{-1}(v) \subseteq {\cal X}_{\cal A} \times \{ x_{{\cal A}^c}\}$ for
some ${\cal A} \in \overline{{\cal L}}$ and $x_{{\cal A}^c} \in {\cal X}_{{\cal A}^c}$, i.e.,
the value of $X_{{\cal A}^c}$ is determined from $V=v$. Thus, $P_{X_{\cal L} V}$
factorizes as in \eqref{eq:proof-symbol-wise-iid-factorize}. 

On the other hand, since function $f$ satisfies Condition \ref{condition:new-2}, for each ${\cal A} \in \overline{{\cal L}}$
and any $a_{\cal A} \in {\cal X}_{{\cal A}}$, the value $[a_{\cal A}]_{\overline{{\cal X}}_{\cal A}}$ is uniquely
determined from the list $( f(a_{\cal A}, b_{{\cal A}^c}) : b_{{\cal A}^c} \in {\cal X}_{{\cal A}^c})$.
Thus, we can construct a mapping satisfying Condition \ref{condition:informative-1} of Definition \ref{definition:informative-function}.
Thus, the symbol-wise function $f^n$ is semi $\overline{{\cal X}}_{\cal A}$-informative for every ${\cal A} \in \overline{{\cal L}}$. \qed 

\section{More Than Two Terminals with Smooth Sources}
\label{section:multi-terminal-smooth}


In Section \ref{section:multi-terminal-iid}, we considered whether ${\cal R}(\bm{X}_{\cal L}|\bm{f}) = {\cal R}_{\san{SW}}(\bm{X}_{\cal L})$
holds or not for the class of i.i.d. sources with positivity condition. In this section, we consider the same problem for the class of smooth
sources. 
In contrast to the classification problem for the class of i.i.d. sources in Section \ref{section:multi-terminal-iid},
we can completely solve the classification problem for the class of smooth sources in this section. 
Let ${\cal F}_{\san{SW}}^{\san{smt}}$ be the set of all functions $f$ such that 
the symbol-wise function induced by $f$ satisfies ${\cal R}(\bm{X}_{\cal L}|\bm{f}) = {\cal R}_{\san{SW}}(\bm{X}_{\cal L})$ 
for any smooth sources. 
Since the class of smooth sources is broader than the class of i.i.d. sources with positivity condition, apparently we have ${\cal F}_{\san{SW}}^{\san{smt}} \subseteq {\cal F}_{\san{SW}}^{\san{iid}}$.
Throughout this section, we use the notations introduced in Section \ref{section:multi-terminal-iid}
such as Definition \ref{definition:projection} and Definition \ref{definition:span}.

Since the class of smooth sources include i.i.d. sources with positivity condition, the necessary condition
in Proposition \ref{propositioon:necessary-HK} is valid for the classification problem for the class of smooth sources.
In fact, since the class of smooth sources includes sources with memory, we can apply the necessary condition
in Proposition \ref{propositioon:necessary-HK} for functions on extended alphabets, which will be used in the proof of
``only if" part of Theorem \ref{theorem:dichotomy-smooth-multi}. 

For the class of smooth sources, 
we can rephrase the conditions in Corollary \ref{corollary:recursion-SW} in a similar manner as Theorem \ref{theorem:condition-symbol-wise-function-iid}.
However, for the class of smooth sources, we can show the matching necessary and sufficient condition in more compact manner.
For that purpose, let us introduce the concept of pseudo identity function,
which is recursively defined as follows.

\begin{definition}[Pseudo Identity] \label{definition:pseudo-identity-smooth}
For a given subset ${\cal A} \subseteq {\cal L}$, the projected function $f_{\cal A}: {\cal X}_{\cal A} \to {\cal V}^{|{\cal X}_{{\cal A}^c}|}$ is said to be
{\em pseudo identity} if either of the following conditions is satisfied:
\begin{enumerate}
\item \label{condition:pseudo-identity-1}
$f_{\cal A}$ is injective; 

\item \label{condition:pseudo-identity-2}
It holds that 
\begin{align} \label{eq:pseudo-identity-smooth-condition}
\tilde{{\cal A}} := \bigcup_{\bm{v} \in {\cal V}^{|{\cal X}_{{\cal A}^c}|}} \mathtt{span} f_{\cal A}^{-1}(\bm{v}) \subsetneq {\cal A},
\end{align}
and, for every ${\cal B} \in \{ \tilde{{\cal A}}, \{ \ell \} : \ell \in {\cal A} \backslash \tilde{{\cal A}} \}$, projected function
$f_{{\cal B}} : {\cal X}_{{\cal B}} \to {\cal V}^{|{\cal X}_{{\cal B}^c}|}$ is pseudo identity.
\end{enumerate}
\end{definition}

\textchange{For the two terminal case, the pseudo identity function is equivalent to the totally sensitive function in \cite{KuzWat15}.}

\begin{remark} \label{remark:pseudo-identity}
Since \eqref{eq:pseudo-identity-smooth-condition} implies that the values of $x_\ell$ for $\ell \in {\cal A}\backslash \tilde{{\cal A}}$
are uniquely determined from $f_{{\cal A}}(x_{{\cal A}})$, $f_{\{\ell \}}$ for $\ell \in {\cal A}\backslash \tilde{{\cal A}}$ are injective, i.e., pseudo identity.
Thus, we only need to verify if $f_{\tilde{{\cal A}}}$ is pseudo identity in the recursion step.
\end{remark}

The following theorem, which will be proved in Sections \ref{proof:theorem-smooth-multi-if} and \ref{proof:theorem-smooth-multi-only-if},
completely solve the classification problem for the class of smooth sources.  

\begin{theorem} \label{theorem:dichotomy-smooth-multi}
A symbol-wise function $f:{\cal X}_{\cal L} \to {\cal V}$ belongs to ${\cal F}_{\san{SW}}^{\san{smt}}$
if and only if $f$ is pseudo identity.
\end{theorem}

Now, we shall illustrate Theorem \ref{theorem:dichotomy-smooth-multi} by some examples.

\begin{example}
The functions in Figs.~\ref{Fig:pattern-three-terminal}(b), (c), (d), (e), and (g)
are pseudo identity, and thus are included in ${\cal F}_{\san{SW}}^{\san{smt}}$.
Since it is not difficult to verify that (b), (c), (d), and (e) are pseudo identity, we only verify (g).
Since 
\begin{align}
\bigcup_{v \in {\cal V}} \mathtt{span} f^{-1}(v) = \{1,2\} \subsetneq \{1,2,3\},
\end{align} 
it suffices to verify $f_{\{1,2\}}$ is pseudo identity (cf.~Remark \ref{remark:pseudo-identity}).
Since 
\begin{align}
\bigcup_{\bm{v} \in {\cal V}^{|{\cal X}_3|}} \mathtt{span} f_{\{1,2\}}^{-1}(\bm{v}) = \{1\} \subsetneq \{1,2\},
\end{align} 
it suffices to verify $f_{\{1\}}$ is pseudo identity; it is pseudo identity since it is injective. 
\end{example}

\begin{example}
The functions in Figs.~\ref{Fig:pattern-three-terminal}(a), (f), (h), (i), and (j) are not pseudo identity,
and thus are not included in ${\cal F}_{\san{SW}}^{\san{smt}}$. For (a), (f), (h), (j), since 
\begin{align}
\bigcup_{v \in {\cal V}} \mathtt{span} f^{-1}(v) = \{1,2,3\},
\end{align}
those functions violate the conditions of pseudo identity. On the other hand, for (i), even though
\begin{align}
\bigcup_{v \in {\cal V}} \mathtt{span} f^{-1}(v) = \{1,2\} \subsetneq \{1,2,3\},
\end{align}
the projected function $f_{\{1,2\}}$ is not pseudo identity since
\begin{align}
\bigcup_{\bm{v} \in {\cal V}^{|{\cal X}_3|}} \mathtt{span} f_{\{1,2\}}^{-1}(\bm{v}) = \{1,2\}.
\end{align}
Note that ${\cal F}_{\san{SW}}^{\san{smt}} \subsetneq {\cal F}_{\san{SW}}^{\san{iid}}$. In fact, 
the functions in (a) and (h) are included in ${\cal F}_{\san{SW}}^{\san{iid}} \backslash {\cal F}_{\san{SW}}^{\san{smt}}$.
Although it is shown that the function in (f) does not belong to ${\cal F}_{\san{SW}}^{\san{smt}}$, it is not clear whether 
it is included in  ${\cal F}_{\san{SW}}^{\san{iid}} \backslash {\cal F}_{\san{SW}}^{\san{smt}}$ or not
(see also Example \ref{example:unclear-case}).
\end{example}

\begin{example}
The function in Fig.~\ref{Fig:pattern-recursion-3}(a) is not pseudo identity. In fact, even though
\begin{align}
\bigcup_{v \in {\cal V}} \mathtt{span} f^{-1}(v) = \{2,3\} \subsetneq \{1,2,3\},
\end{align}
$f_{\{2,3\}}$ is not pseudo identity since 
\begin{align}
\bigcup_{\bm{v} \in {\cal V}^{|{\cal X}_1|}} f_{\{2,3\}}^{-1}(\bm{v}) = \{2,3\}.
\end{align}
\end{example}

Finally, let us consider an example of function such that multiple rounds of recursion are needed to verify
that is pseudo identity.

\begin{example} \label{example:multiple-recursion}
We define function $f$ recursively as follows, where ${\cal X}_\ell = \{0,1\}$ for every $\ell \in {\cal L}$.
Let $f_{[1]}(x_1) := x_1$. Then, for $\ell = 2,\ldots,L$, let 
\begin{align}
f_{[\ell]}(x_1,\ldots,x_{\ell-1},0) &:= f_{[\ell-1]}(x_1,\ldots,x_{\ell-1}), \\
f_{[\ell]}(x_1,\ldots,x_{\ell-1},1) &:= \ell. 
\end{align}
Then, the function $f = f_{[L]} : {\cal X}_{\cal L} \to {\cal V} = \{0,1,\ldots,L\}$ is pseudo identity. 
In fact, even though $f$ is not injective, we can verify that 
\begin{align}
\bigcup_{v \in {\cal V}} \mathtt{span} f^{-1}(v) = {\cal L}\backslash \{L\} \subsetneq {\cal L}.
\end{align}
Similarly, even though $f_{{\cal L}\backslash \{\ell+1,\ldots,L\}}$ is not injective for $\ell =L-1,\ldots,2$, we can verify that 
\begin{align}
\bigcup_{\bm{v} \in {\cal V}^{2^{L-\ell}}} \mathtt{span} f_{{\cal L}\backslash \{\ell+1,\ldots,L\}}^{-1}(\bm{v}) = \{1,\ldots,\ell-1\} \subsetneq \{1,\ldots,\ell\}.
\end{align}
Eventually, we can verify that $f_{\{1\}}$ is injective. The special case for $L = 3$ is the function in Fig.~\ref{Fig:pattern-three-terminal}(g).
\end{example}

It should be emphasized that the function in Example \ref{example:multiple-recursion} belongs to ${\cal F}_{\san{SW}}^{\san{smt}}$
even though the image size $|{\cal V}| = L+1$ of the function is much smaller (exponential in the number of terminals)
than the input size $|{\cal X}_{\cal L}| = 2^L$.

\subsection{Proof of ``if" part of Theorem \ref{theorem:dichotomy-smooth-multi}} \label{proof:theorem-smooth-multi-if}

If $f$ is injective, then it is trivial that $f \in {\cal F}_{\san{SW}}^{\san{smt}}$; thus, we 
assume that $f$ is not injective. Then, since $f$ is pseudo identity, 
there exists a sequence of subsets 
\begin{align}
{\cal A}_k \subsetneq {\cal A}_{k-1} \subsetneq \cdots \subsetneq {\cal A}_1 \subsetneq {\cal A}_0 = {\cal L}
\end{align}
for some $k \ge 1$ such that \eqref{eq:pseudo-identity-smooth-condition} holds with ${\cal A}= {\cal A}_{i-1}$
and $\tilde{{\cal A}} = {\cal A}_i$ for $i=1,\ldots,k$, and $f_{{\cal A}_k}$ is injective. 
From this $\{ {\cal A}_i \}_{i=1}^k$, we shall construct a sequence of partitions $\{ \overline{{\cal L}}^{(i)} \}_{i=1}^k$
and a sequence of tuples of partitions $\{ \overline{{\cal X}}_{\cal L}^{(i)} \}_{i=1}^k$ satisfying the 
requirement of Corollary \ref{corollary:recursion-SW}. In fact, we set
\begin{align}
\overline{{\cal L}}^{(i)} = \{ {\cal A}_i, \{ \ell \} : \ell \in {\cal L}\backslash {\cal A}_i \}
\end{align}
for $i=1,\ldots,k$; 
\begin{align} \label{eq:definition-partiion-k-1}
\overline{{\cal X}}^{(i)}_\ell \equiv {\cal X}_\ell \mbox{ for } \ell \in {\cal L}\backslash {\cal A}_{i+1} \mbox{ and } \overline{{\cal X}}^{(i)}_\ell = \{ {\cal X}_\ell \} \mbox{ for } \ell \in {\cal A}_{i+1}
\end{align}
for $i=1,\ldots,k-1$; and 
\begin{align} \label{eq:definition-partiion-k}
\overline{{\cal X}}^{(k)}_\ell \equiv {\cal X}_\ell \mbox{ for } \ell \in {\cal L}.
\end{align}
Then, we can verify the requirement of Corollary \ref{corollary:recursion-SW},
i.e., $(\bm{f}, \bm{f}_{\overline{{\cal X}}_{\cal L}^{(1)}},\ldots, \bm{f}_{\overline{{\cal X}}_{\cal L}^{(i-1)}})$
induces conditional independence for $(\bm{X}_{\cal L}, \overline{{\cal L}}^{(i)})$ and 
semi $\overline{{\cal X}}_{\cal A}^{(i)}$-informative for every ${\cal A} \in \overline{{\cal L}}^{(i)}$
for each $i=1,\ldots,k$, as follows.

For $i=1$, since 
\begin{align}
{\cal A}_1 = \bigcup_{v \in {\cal V}} \mathtt{span} f^{-1}(v),
\end{align}
the values of $x_\ell$ for $\ell \in {\cal L}\backslash {\cal A}_1$ are uniquely determined from $f(x_{\cal L})$.
Thus, $\bm{f}$ induces conditional independence for $(\bm{X}_{\cal L}, \overline{{\cal L}}^{(1)})$.
Furthermore, for $\ell \in {\cal L}\backslash{\cal A}_1$, $\bm{f}$ is $\overline{{\cal X}}_\ell^{(1)}$-informative.\footnote{Note that
$\overline{{\cal X}}_\ell^{(1)}$ for $\ell \in {\cal L}\backslash{\cal A}_1$ is defined by \eqref{eq:definition-partiion-k-1} when $k \ge 2$ and
defined by \eqref{eq:definition-partiion-k} when $k =1$.}
If $k=1$, then $\bm{f}$ is semi $\overline{{\cal X}}_{{\cal A}_1}^{(1)}$-informative since $f_{{\cal A}_1}$ is injective.
If $k \ge 2$, since
\begin{align}
{\cal A}_{2} = \bigcup_{\bm{v} \in {\cal V}^{|{\cal X}_{{\cal A}_{1}^c}|}} \mathtt{span} f_{{\cal A}_{1}}^{-1}(\bm{v}),
\end{align}
the values of $x_\ell$ for $\ell \in {\cal A}_2\backslash {\cal A}_1$ are uniquely determined from $f_{{\cal A}_1}(x_{{\cal A}_1})$.
Furthermore, $\overline{{\cal X}}_\ell^{(1)} = \{ {\cal X}_\ell \}$ is
the trivial partition for $\ell \in {\cal A}_1$.
Thus, $\bm{f}$ is $\overline{{\cal X}}_{{\cal A}_1}^{(1)}$-informative.

For $i\ge 2$, since $\overline{{\cal X}}_\ell^{(i-1)} \equiv {\cal X}_\ell$ for $\ell \in {\cal L}\backslash {\cal A}_{i}$,
the values of $x_\ell$ is trivially determined from local function $f_{\overline{{\cal X}}_{{\cal L}}^{(i-1)}}(x_{\cal L})$
for $\ell \in {\cal L}\backslash {\cal A}_{i}$.
Thus, $(\bm{f}, \bm{f}_{\overline{{\cal X}}_{\cal L}^{(1)}},\ldots, \bm{f}_{\overline{{\cal X}}_{\cal L}^{(i-1)}})$ induces conditional independence
for $(\bm{X}_{\cal L}, \overline{{\cal L}}^{(i)})$. Furthermore, for $\ell \in {\cal L}\backslash {\cal A}_{i}$, 
$(\bm{f}, \bm{f}_{\overline{{\cal X}}_{\cal L}^{(1)}},\ldots, \bm{f}_{\overline{{\cal X}}_{\cal L}^{(i-1)}})$ is semi $\overline{{\cal X}}_\ell^{(i)}$-informative\footnote{Note
that $\overline{{\cal X}}_\ell^{(i)}$ for $\ell \in {\cal L}\backslash {\cal A}_{i}$ is defined by \eqref{eq:definition-partiion-k-1} when $i \le k-1$ and
defined by \eqref{eq:definition-partiion-k} when $i=k$.}.
If $i = k$, then $(\bm{f}, \bm{f}_{\overline{{\cal X}}_{\cal L}^{(1)}},\ldots, \bm{f}_{\overline{{\cal X}}_{\cal L}^{(i-1)}})$
is semi $\overline{{\cal X}}_{{\cal A}_{i}}^{(i)}$-informative since $f_{{\cal A}_{i}}$ is injective.
If $i \le k-1$, since 
\begin{align}
{\cal A}_{i+1} = \bigcup_{\bm{v} \in {\cal V}^{|{\cal X}_{{\cal A}_{i}^c}|}} \mathtt{span} f_{{\cal A}_{i}}^{-1}(\bm{v}),
\end{align}
the values of $x_\ell$ for $\ell \in {\cal A}_{i}\backslash {\cal A}_{i+1}$ are uniquely determined from $f_{{\cal A}_{i}}(x_{{\cal A}_{i}})$.
Furthermore, $\overline{{\cal X}}_\ell^{(i)} = \{ {\cal X}_\ell \}$ is the trivial partition for $\ell \in {\cal A}_{i+1}$.
Thus, $(\bm{f}, \bm{f}_{\overline{{\cal X}}_{\cal L}^{(1)}},\ldots, \bm{f}_{\overline{{\cal X}}_{\cal L}^{(i-1)}})$ is 
$\overline{{\cal X}}_{{\cal A}_i}^{(i)}$-informative.  \qed

\subsection{Proof of ``only if" part of Theorem \ref{theorem:dichotomy-smooth-multi}} \label{proof:theorem-smooth-multi-only-if}

Since $f$ is not pseudo identity (cf.~Definition \ref{definition:pseudo-identity-smooth}), either 
\begin{enumerate}
\renewcommand\labelenumi{(\roman{enumi})}
\renewcommand\theenumi\labelenumi
\item \label{smooth-violation-condition-1}
there exists $\ell \in {\cal L}$ such that $f_{\{\ell \}}$ is not injective;

\item \label{smooth-violation-condition-2}
there exists ${\cal A} \subseteq {\cal L}$ such that 
\begin{align}
\bigcup_{\bm{v} \in {\cal V}^{|{\cal X}_{{\cal A}^c}|}} \mathtt{span} f_{{\cal A}}^{-1}(\bm{v}) = {\cal A}.
\end{align}
\end{enumerate}
We construct a function such that ${\cal R}_{\san{SW}}(\bm{X}_{\cal L}) \subsetneq {\cal R}(\bm{X}_{\cal L}|\bm{f})$
in a similar manner as the proof of ``only if part" of two terminal dichotomy theorem in \cite[Theorem 3]{KuzWat15}.

When Case \ref{smooth-violation-condition-1} is true, the function $f$ violates the necessary condition in Proposition \ref{propositioon:necessary-HK}.
Thus, there exists an i.i.d. source $\bm{X}_{\cal L}$ such that the function computation region ${\cal R}(\bm{X}_{\cal L}|\bm{f})$
is strictly broader than the Slepian-Wolf region ${\cal R}_{\san{SW}}(\bm{X}_{\cal L})$.

When Case \ref{smooth-violation-condition-2} is true, for each $\ell \in {\cal A}$, there exist
$a_{\cal A}^{(\ell)}, \hat{a}_{\cal A}^{(\ell)} \in {\cal X}_{\cal A}$ such that 
\begin{align} \label{eq:proof-smooth-necessary-violation-1}
a_\ell^{(\ell)} \neq \hat{a}_\ell^{(\ell)}
\end{align}
and
\begin{align} \label{eq:proof-smooth-necessary-violation-2}
f_{\cal A}(a_{\cal A}^{(\ell)}) = f_{\cal A}(\hat{a}_{\cal A}^{(\ell)}).
\end{align}
By denoting $m = |{\cal A}|$, let us consider $m$-length copy of function $f$, i.e.,
\begin{align}
f^m(x^m_{\cal L}) := (f(x_{{\cal L},1}),\ldots, f(x_{{\cal L},m}))
\end{align}
for $x_{\cal L}^m = (x_{{\cal L},1},\ldots,x_{{\cal L},m})$. Then, this function $f^m$ violates
the necessary condition in Proposition \ref{propositioon:necessary-HK} as a function from ${\cal X}_{\cal L}^m$ to ${\cal V}^m$.
In fact, because of \eqref{eq:proof-smooth-necessary-violation-1} and \eqref{eq:proof-smooth-necessary-violation-2}, 
the pair $(a_{\cal A}^{(1)},\ldots,a_{\cal A}^{(m)})$ and $(\hat{a}_{\cal A}^{(1)},\ldots,\hat{a}_{\cal A}^{(m)})$ violate the requirement of the necessary condition. 
Thus, there exists a block i.i.d. source 
$\tilde{X}_{\cal L}^{m,N} = (\tilde{X}_{\cal L}^m[1],\ldots,\tilde{X}_{\cal L}^m[N])$ on extended alphabet ${\cal X}_{\cal L}^m$ such that 
$\tilde{\bm{X}}_{\cal L}^m = \{ \tilde{X}_{\cal L}^{m,N} \}_{N=1}^\infty$ and $\bm{f}^m = \{ (f^m)^N \}_{N=1}^\infty$ satisfy
\begin{align} \label{eq:improvement-extended-alphabet}
{\cal R}_{\san{SW}}(\tilde{\bm{X}}_{\cal L}^m) \subsetneq {\cal R}(\tilde{\bm{X}}_{\cal L}^m | \bm{f}^m).
\end{align}
Now, let $\bm{X} = \tilde{\bm{X}}^m_{\cal L}$, which is a smooth source on ${\cal X}_{\cal L}$,\footnote{When $n \neq m N$
for any integer $N$, we appropriately pad some random variables, which does not affect the rate regions asymptotically.} 
and $\bm{f} = \bm{f}^m$. Then, \eqref{eq:improvement-extended-alphabet} implies (with scaling by the common factor $\frac{1}{m}$)
\begin{align}
{\cal R}_{\san{SW}}(\bm{X}_{\cal L}) \subsetneq {\cal R}(\bm{X}_{\cal L}|\bm{f}).
\end{align}
\qed


\section{Conclusion} \label{section:conclusion}

In this paper, we developed a general approach to derive converse bounds
on multiterminal distributed function computing. As an application of the proposed approach, we considered the function classification 
problem for the class of i.i.d. sources in multiterminal setting, and derived a novel sufficient condition that strictly 
subsume the sufficient conditions derived by Han-Kobayashi. As we have seen in Example \ref{example:unclear-case}, 
there are some functions that cannot be classified by our approach. Currently, it is not clear whether 
the sufficient condition is loose or the necessary condition is loose.
If the necessary condition could be improved, a novel achievability scheme that is more sophisticated than
the K\"orner-Marton type scheme used in \cite{HanKob87} may be needed; possibly, the approaches studied in the literature \cite{KriPra:11, HuaSko:17}
might be useful.

\appendix 

\subsection{Technical Lemmas}

The following lemma is a simple property of the variational distance
(e.g.~see \cite[Lemma 11.3]{mitzenmacher:book}).

\begin{lemma} \label{lemma:property-of-variational-distance}
For a pair of random variables $(X,Y)$ on the same alphabet, we have
\begin{align}
\| P_X - P_Y \|_1 \le \Pr( X \neq Y).
\end{align}
\end{lemma}

The following lemma is an immediate consequence of 
the alternative definition of the variational distance.
\begin{lemma} \label{lemma:probability-variational-distance}
For a pair of random variables $(X,Y)$ on the same alphabet and any event ${\cal E}$, we have
\begin{align}
\Pr( X \in {\cal E}) \le \Pr(Y \in {\cal E}) + \| P_X - P_Y \|_1.
\end{align}
\end{lemma}

In the converse proof of Theorem \ref{theorem:general}, we need to boost a code
with small symbol error probability to a code with small block error probability.
For that purpose, we use the following lemma shown in \cite[Lemma 4]{KuzWat16}.
For $0 < \beta < 1/2$ and an integer $m$, let 
\begin{align}
\nu_n(\beta,m) &:= \sum_{i=0}^{\lceil n \beta \rceil -1} (m-1)^i {n \choose i} \le n m^{n\beta} 2^{n h(\beta)}.
\end{align}

\begin{lemma}\cite{KuzWat16} \label{lemma:boosting}
Suppose that $(W^n,X^n)$ on ${\cal X}^n \times {\cal X}^n$ satisfies 
\begin{align}
\Pr\left( \frac{1}{n} d_H(W^n,X^n)) \ge \beta \right) \le \varepsilon_n.
\end{align}
Then, there exists encoders $\kappa_n:{\cal X}^n \to {\cal K}_n$ 
with $|{\cal K}_n| \le 2^{n\delta}$ and a decoder $\tau_n:{\cal K}_n \times {\cal X}^n \to {\cal X}^n$ such that 
\begin{align}
\Pr\left( \tau_n(\kappa_n(X^n), W^n) \neq X^n \right) \le \varepsilon_n + \nu_n(\beta,|{\cal X}|) 2^{-n\delta}.
\end{align}
\end{lemma}

\subsection{Proof of \eqref{eq:smooth-bound}} \label{appendix:smooth-bound}

The proof is essentially the same as \cite[Proof of Theorem 1]{KuzWat16} (see also \cite[Sec. III]{KuzWat16} 
for the high level idea of the argument). 
For convenience of notation, we identify ${\cal X}_{{\cal A}^c}$ as $\{0,1,\ldots, |{\cal X}_{{\cal A}^c}|-1\}$.
Let $\pi_i : {\cal X}_{{\cal A}^c}^n \to {\cal X}_{{\cal A}^c}^n$ be the permutation that shift only
$i$th symbol of $\bm{x}_{{\cal A}^c} \in {\cal X}_{{\cal A}^c}^n$, i.e., $x_{{\cal A}^c,i} \mapsto x_{{\cal A}^c,i}+1 \pmod{|{\cal X}_{{\cal A}^c}|}$.
By using this cyclic permutation $\pi_i$, we can rewrite \eqref{eq:smooth-bound} as follows:
\begin{align}
\lefteqn{
\Pr\left(  \left( \psi_n\big( \varphi_n^{({\cal A})}(X_{\cal A}^n), \varphi_n^{({\cal A}^c)}( a_{{\cal A}^c} X_{{\cal A}^c}^{(-i)}) \big) : a_{{\cal A}^c} \in {\cal X}_{{\cal A}^c} \right) 
 \neq \left( f_n(X_{\cal A}^n, a_{{\cal A}^c} X_{{\cal A}^c}^{(-i)}) : a_{{\cal A}^c} \in {\cal X}_{{\cal A}^c} \right) \right)
} \\
&= \Pr\left(  \exists a_{{\cal A}^c} \in {\cal X}_{{\cal A}^c} \mbox{ s.t. }
  \psi_n\big( \varphi_n^{({\cal A})}(X_{\cal A}^n), \varphi_n^{({\cal A}^c)}( a_{{\cal A}^c} X_{{\cal A}^c}^{(-i)}) \big)  
 \neq f_n(X_{\cal A}^n, a_{{\cal A}^c} X_{{\cal A}^c}^{(-i)})  \right) \\
&= \sum_{\bm{x}_{{\cal A}},\bm{x}_{{\cal A}^c}} P_{X^n_{{\cal L}}}(\bm{x}_{{\cal A}}, \bm{x}_{{\cal A}^c})
 \bol{1}\bigg[
  \exists a_{{\cal A}^c} \in {\cal X}_{{\cal A}^c} \mbox{ s.t. }
  \psi_n\big( \varphi_n^{({\cal A})}(\bm{x}_{\cal A}), \varphi_n^{({\cal A}^c)}( a_{{\cal A}^c} \bm{x}_{{\cal A}^c}^{(-i)}) \big)  
 \neq f_n(\bm{x}_{\cal A}, a_{{\cal A}^c} \bm{x}_{{\cal A}^c}^{(-i)})
 \bigg]  \\
&= \sum_{\bm{x}_{{\cal A}},\bm{x}_{{\cal A}^c}} P_{X^n_{{\cal L}}}(\bm{x}_{{\cal A}}, \bm{x}_{{\cal A}^c})
 \bol{1}\bigg[
  \exists b \in {\cal X}_{{\cal A}^c} \mbox{ s.t. }
  \psi_n\big( \varphi_n^{({\cal A})}(\bm{x}_{\cal A}), \varphi_n^{({\cal A}^c)}( \pi_i^b( \bm{x}_{{\cal A}^c})) \big)  
 \neq f_n(\bm{x}_{\cal A}, \pi_i^b( \bm{x}_{{\cal A}^c}))
 \bigg] \\ 
&\le \sum_{b \in {\cal X}_{{\cal A}^c}} \sum_{\bm{x}_{{\cal A}},\bm{x}_{{\cal A}^c}} P_{X^n_{{\cal L}}}(\bm{x}_{{\cal A}}, \bm{x}_{{\cal A}^c})
 \bol{1}\bigg[
  \psi_n\big( \varphi_n^{({\cal A})}(\bm{x}_{\cal A}), \varphi_n^{({\cal A}^c)}( \pi_i^b( \bm{x}_{{\cal A}^c})) \big)  
 \neq f_n(\bm{x}_{\cal A}, \pi_i^b( \bm{x}_{{\cal A}^c}))
 \bigg], \label{eq:cyclic-version}
\end{align}
where $\pi_i^b$ means $b$ times application of $\pi_i$. 
Then, since $\bm{X}_{{\cal L}}$ is smooth, for every $b \in {\cal X}_{{\cal A}^c}$ and $i \in [1:n]$, we can bound 
each term of \eqref{eq:cyclic-version} as follows:
\begin{align}
\lefteqn{ \sum_{\bm{x}_{{\cal A}},\bm{x}_{{\cal A}^c}} P_{X^n_{{\cal L}}}(\bm{x}_{{\cal A}}, \bm{x}_{{\cal A}^c})
 \bol{1}\bigg[
  \psi_n\big( \varphi_n^{({\cal A})}(\bm{x}_{\cal A}), \varphi_n^{({\cal A}^c)}( \pi_i^b( \bm{x}_{{\cal A}^c})) \big)  
 \neq f_n(\bm{x}_{\cal A}, \pi_i^b( \bm{x}_{{\cal A}^c}))
 \bigg] } \\
&\le \sum_{\bm{x}_{{\cal A}},\bm{x}_{{\cal A}^c}} \frac{1}{q} P_{X^n_{{\cal L}}}(\bm{x}_{{\cal A}}, \pi_i^b(\bm{x}_{{\cal A}^c}))
 \bol{1}\bigg[
  \psi_n\big( \varphi_n^{({\cal A})}(\bm{x}_{\cal A}), \varphi_n^{({\cal A}^c)}( \pi_i^b( \bm{x}_{{\cal A}^c})) \big)  
 \neq f_n(\bm{x}_{\cal A}, \pi_i^b( \bm{x}_{{\cal A}^c}))
 \bigg] \\
&= \frac{1}{q} \Pr\bigg( \psi_n\big( \varphi_n^{({\cal A})}(X_{\cal A}^n), \varphi_n^{({\cal A}^c)}( X_{{\cal A}^c}^n) \big)
 \neq  f_n(X_{\cal A}^n, X_{{\cal A}^c}^n) \bigg) \\
&\le \frac{\varepsilon_n}{q}.
\end{align}
Substituting this bound into \eqref{eq:cyclic-version}, we have the desired bound. \qed

\section*{Acknowledgements} 

The author would like to thank Shigeaki Kuzuoka for helpful comments.


\bibliographystyle{./IEEEtranS}
\bibliography{../../../09-04-17-bibtex/reference}

\end{document}